

\input harvmac

\noblackbox
\baselineskip 20pt plus 2pt minus 2pt

\overfullrule=0pt



\def\bs{\bigskip}

\def\hb{\hfill\break}
\def\qq{\qquad}
\def\bl{\bigl}
\def\br{\bigr}

\def\IR{\relax{\rm I\kern-.18em R}}

\def\np {  Nucl. Phys. }


\def\a{\alpha}
\def\A{\Alpha}
\def\b{\beta}

\def\g{\gamma}
\def\G{\Gamma}
\def\d{\delta}

\def\e{\epsilon}

\def\m{\mu}
\def\n{\nu}

\def\Om{\Omega}

\def\s{\sigma}

\def\cM{{\cal M}}

\def\cV{{\cal V}}
\def\ctV{\tilde{\cal V}}
\def\cL{{\cal L}}
\def\cR{{\cal R}}

\def\IR{\relax{\rm I\kern-.18em R}}

\def\ghc{ G^c_h }

\def \bd {\bar \del}
\def \bh { {\bar h} }

\def \z { {\bar z} }
\def \A { {\bar A} }
\def \tA  { {\tilde {A }}}

\def \ha {{1\over 2}}

\def \ov {\over}



\lref\BSthree{I. Bars and K. Sfetsos, Mod. Phys. Lett. {\bf A7} (1992) 1091.}

\lref\BShet{I. Bars and K. Sfetsos, Phys. Lett. {\bf 277B} (1992) 269.}

\lref\BSglo{I. Bars and K. Sfetsos, Phys. Rev. {\bf D46} (1992) 4495.}

\lref\BSexa{I. Bars and K. Sfetsos, Phys. Rev. {\bf D46} (1992) 4510.}

\lref\SFET{K. Sfetsos, Nucl. Phys. {\bf B389} (1993) 424.}

\lref\BSslsu{I. Bars and K. Sfetsos, Phys. Lett. {\bf 301B} (1993) 183.}

\lref\BSeaction{I. Bars and K. Sfetsos, Phys. Rev. {\bf D48} (1993) 844.}

\lref\BN{ I. Bars and D. Nemeschansky, Nucl. Phys. {\bf B348} (1991) 89.}

\lref\WIT{E. Witten, Phys. Rev. {\bf D44} (1991) 314.}

 \lref\IBhet{ I. Bars, Nucl. Phys. {\bf B334} (1990) 125. }

 \lref\IBCS{ I. Bars, ``String Propagation on Black Holes'', USC-91-HEP-B3.\hb
{\it Curved Space-time Strings and Black Holes},
in Proc.
 {\it XX$^{th}$ Int. Conf. on Diff. Geometrical Methods in Physics}, eds. S.
 Catto and A. Rocha, Vol. 2, p. 695, (World Scientific, 1992).}

 \lref\CRE{M. Crescimanno, Mod. Phys. Lett. {\bf A7} (1992) 489.}

\lref\clapara{K. Bardakci, M. Crescimanno and E. Rabinovici,
Nucl. Phys. {\bf B344} (1990) 344.}

\lref\MSW{G. Mandal, A. Sengupta and S. Wadia,
Mod. Phys. Lett. {\bf A6} (1991) 1685.}

 \lref\HOHO{J. B. Horne and G.T. Horowitz, Nucl. Phys. {\bf B368} (1992) 444.}

 \lref\FRA{E. S. Fradkin and V. Ya. Linetsky, Phys. Lett. {\bf 277B}
          (1992) 73.}

 \lref\ISH{N. Ishibashi, M. Li, and A. R. Steif,
         Phys. Rev. Lett. {\bf 67} (1991) 3336.}

 \lref\HOR{P. Horava, Phys. Lett. {\bf 278B} (1992) 101.}

 \lref\RAI{E. Raiten, ``Perturbations of a Stringy Black Hole'',
         Fermilab-Pub 91-338-T.}

 \lref\GER{D. Gershon, ``Exact Solutions of Four-Dimensional Black Holes in
         String Theory'', TAUP-1937-91.}

 \lref \GIN {P. Ginsparg and F. Quevedo,  Nucl. Phys. {\bf B385} (1992) 527. }

 \lref\HOHOS{ J. H. Horne, G.T. Horowitz and A. R. Steif, Phys. Rev. Lett.
 {\bf 68} (1991) 568.}

 \lref\groups{
 M. Crescimanno. Mod. Phys. Lett. {\bf A7} (1992) 489. \hb
 J. B. Horne and G.T. Horowitz, Nucl. Phys. {\bf B368} (1992) 444. \hb
 E. S. Fradkin and V. Ya. Linetsky, Phys. Lett. {\bf 277B} (1992) 73. \hb
 P. Horava, Phys. Lett. {\bf 278B} (1992) 101.\hb
 E. Raiten, ``Perturbations of a Stringy Black Hole'',
         Fermilab-Pub 91-338-T.\hb
 D. Gershon, ``Exact Solutions of Four-Dimensional Black Holes in
         String Theory'', TAUP-1937-91.}

\lref\NAWIT{C. Nappi and E. Witten, Phys. Lett. {\bf 293B} (1992) 309.}

\lref\FRATSE{E. S. Fradkin and A. A. Tseytlin,
Phys. Lett. {\bf 158B} (1985) 316.}

\lref\CALLAN{ C. G. Callan, D. Friedan, E. J. Martinec and M. Perry,
Nucl. Phys. {\bf B262} (1985) 593.}

\lref\DB{L. Dixon, J. Lykken and M. Peskin, Nucl. Phys.
{\bf B325} (1989) 325.}

\lref\IB{I. Bars, Nucl. Phys. {\bf B334} (1990) 125.}

\lref\BUSCHER{T. Buscher, Phys. Lett. {\bf 194B} (1087) 59;
Phys. Lett. {\bf 201B} (1988) 466.}

\lref\RV{M. Rocek and E. Verlinde, Nucl. Phys. {\bf B373} (1992) 630.}
\lref\GR{A. Giveon and M. Rocek, Nucl. Phys. {\bf B380} (1992) 128. }

\lref\nadual{X. C. de la Ossa and F. Quevedo,
Nucl. Phys. {\bf B403} (1993) 377.}

\lref\duearl{B.E. Fridling and A. Jevicki,
Phys. lett. {\bf B134} (1984) 70.\hb
E.S. Fradkin and A.A. Tseytlin, Ann. Phys. {\bf 162} (1985) 31.}

\lref\SENrev{A. Sen, ``Black Holes and Solitons in String Theory'',
TIFR-TH-92-57.}

\lref\TSEd{A. A. Tseytlin, Mod. Phys. Lett. {\bf A6} (1991) 1721.}

\lref\TSESC{A. S. Schwarz and A. A. Tseytlin, ``Dilaton shift under duality
and torsion of elliptic complex'', IMPERIAL/TP/92-93/01. }

\lref\Dualone{K. Meissner and G. Veneziano,
Phys. Lett. {\bf B267} (1991) 33;
Mod. Phys. Lett. {\bf A6} (1991) 3397. \hb
M. Gasperini and G. Veneziano, Phys. Lett. {\bf 277B} (1992) 256. \hb
M. Gasperini, J. Maharana and G. Veneziano, Phys. Lett. {\bf 296B} (1992) 51.}

\lref\rovr{K. Kikkawa and M. Yamasaki, Phys. Lett. {\bf B149} (1984) 357.\hb
N. Sakai and I. Senda, Prog. theor. Phys. {\bf 75} (1986) 692.}

\lref\narain{K.S. Narain, Phys. Lett. {\bf B169} (1986) 369.\hb
K.S. Narain, M.H. Sarmadi and C. Vafa, Nucl. Phys. {\bf B288} (1987) 551.}

\lref\GV{P. Ginsparg and C. Vafa, Nucl. Phys. {\bf B289} (1987) 414.}
\lref\nssw{V. Nair, A. Shapere, A. Strominger and F. Wilczek,
Nucl. Phys. {\bf B287} (1987) 402.}

\lref\vafa{C. Vafa, ``Strings and Singularities'', HUTP-93/A028.}

\lref\Dualtwo{A. Sen,
Phys. Lett. {\bf B271} (1991) 295;\ ibid. {\bf B274} (1992) 34;
Phys. Rev. Lett. {\bf 69} (1992) 1006. \hb
S. Hassan and A. Sen, Nucl. Phys. {\bf B375} (1992) 103. \hb
J. Maharana and J. H. Schwarz, Nucl. Phys. {\bf B390} (1993) 3.\hb
A. Kumar, Phys. Lett. {\bf B293} (1992) 49.}

\lref\dualmargi{S.F. Hassan and A. Sen, Nucl. Phys. {\bf B405} (1993) 143.\hb
M. Henningson and C. Nappi, Phys. Rev. {\bf D48} (1993) 861.}

\lref\bv{R. Brandenberger and C. Vafa, Nucl. Phys. {\bf B316} (1989) 301.}
\lref\gsvy{B.R. Greene, A. Shapere, C. Vafa and S.T. Yau,
Nucl. Phys. {\bf B337} (1990) 1.}
\lref\tv{A.A. Tseytlin and C. Vafa,  Nucl. Phys. {\bf B372} (1992) 443.}

\lref\grvm{A. Shapere and F. Wilczek, Nucl. Phys. {\bf B320} (1989) 609.\hb
A. Giveon, E. Rabinovici and G. Veneziano,
Nucl. Phys. {\bf B322} (1989) 167.\hb
A. Giveon, N. Malkin and E. Rabinovici, Phys. Lett. {\bf B238} (1990) 57.}

\lref\GRV{M. Gasperini, R. Ricci and G. Veneziano,
Phys. Lett. {\bf B319} (1993) 438.}

\lref\KIRd{E. Kiritsis, Nucl. Phys. {\bf B405} (1993) 109.}

\lref\gpr{A. Giveon, M. Porrati and E. Rabinovici, ``Target Space Duality
in String Theory'', RI-1-94, hepth/9401139.}

\lref\slt{A. Giveon, Mod. Phys. Lett. {\bf A6} (1991) 2843.}

\lref\GIPA{A. Giveon and A. Pasquinucci, ``On cosmological string backgrounds
with toroidal isometries'', IASSNS-HEP-92/55, August 1992.}

\lref\KASU{Y. Kazama and H. Suzuki, Nucl. Phys. {\bf B234} (1989) 232. \hb
Y. Kazama and H. Suzuki Phys. Lett. {\bf 216B} (1989) 112.}

\lref\WITanom{E. Witten, Comm. Math. Phys. {\bf 144} (1992) 189.}

\lref\WITnm{E. Witten, Nucl. Phys. {\bf B371} (1992) 191.}

\lref\IBhetero{I. Bars, Phys. Lett. {\bf 293B} (1992) 315.}

\lref\IBerice{I. Bars, {\it Superstrings on Curved Space-times}, Lecture
delivered at the Int. workshop on {\it String Quantum Gravity and Physics
at the Planck Scale}, Erice, Italy, June 1992.}

\lref\DVV{R. Dijkgraaf, E. Verlinde and H. Verlinde, Nucl. Phys. {\bf B371}
(1992) 269.}

\lref\TSEY{A. A. Tseytlin, Phys. Lett. {\bf 268B} (1991) 175.}

\lref\JJP{I. Jack, D. R. T. Jones and J. Panvel,
          Nucl. Phys. {\bf B393} (1993) 95.}

\lref\BST { I. Bars, K. Sfetsos and A. A. Tseytlin, unpublished. }

\lref\TSEYT{ A. A. Tseytlin, Nucl. Phys. {\bf B399} (1993) 601.}

\lref\TSEYTt{A. A. Tseytlin, Nucl. Phys. {\bf B411} (1993) 509.}

 \lref\SHIF { M. A. Shifman, Nucl. Phys. {\bf B352} (1991) 87.}
\lref\SHIFM { H. Leutwyler and M. A. Shifman, Int. J. Mod. Phys. {\bf
A7} (1992) 795. }

\lref\POLWIG { A. M. Polyakov and P. B. Wiegman, Phys.
Lett. {\bf 141B} (1984) 223.  }

\lref\BCR{K. Bardakci, M. Crescimanno
and E. Rabinovici, Nucl. Phys. {\bf B344} (1990) 344. }

\lref\Wwzw{E. Witten, Commun. Math. Phys. {\bf 92} (1984) 455.}

\lref\GKO{P. Goddard, A. Kent and D. Olive, Phys. Lett. {\bf 152B} (1985) 88.}

\lref\Toda{A. N. Leznov and M. V. Saveliev, Lett. Math. Phys. {\bf 3} (1979)
489. \hb A. N. Leznov and M. V. Saveliev, Comm. Math. Phys. {\bf 74}
(1980) 111.}

\lref\GToda{J. Balog, L. Feh\'er, L. O'Raifeartaigh, P. Forg\'acs and A. Wipf,
Ann. Phys. (New York) {\bf 203} (1990) 76; Phys. Lett. {\bf 244B}
(1990) 435.}

\lref\GWZW{ E. Witten, \np {\bf B223} (1983) 422. \hb
K. Bardakci, E. Rabinovici and B. S\"aring, Nucl. Phys. {\bf B299}
(1988) 157. \hb K. Gawedzki and A. Kupiainen, Phys. Lett. {\bf 215B}
(1988) 119. \hb K. Gawedzki and A. Kupiainen, Nucl. Phys. {\bf B320}
(1989) 625. }

\lref\SCH{ D. Karabali, Q-Han Park, H. J. Schnitzer and Z. Yang,
                   Phys. Lett. {\bf B216} (1989) 307. \hb D. Karabali
and H. J. Schnitzer, Nucl. Phys. {\bf B329} (1990) 649. }

 \lref\KIR{E. Kiritsis, Mod. Phys. Lett. {\bf A6} (1991) 2871. }

\lref\BIR{N. D. Birrell and P. C. W. Davies,
{\it Quantum Fields in Curved Space}, Cambridge University Press.}

\lref\WYB{B. G. Wybourn, {\it Classical Groups for Physicists }
(John Wiley \& sons, 1974).}

\lref\Brinkman{H.W. Brinkmann, Math. Ann. {\bf 94} (1925) 119.}

\lref\SANTA{R. Guven, Phys. Lett. {\bf 191B} (1987) 275.\hb
D. Amati and C. Klimcik, Phys. Lett. {\bf 219B} (1989) 443.\hb
G.T. Horowitz and A. R. Steif, Phys. Rev. Lett. {\bf 64} (1990) 260;
Phys. Rev. {\bf D42} (1990) 1950.\hb
R.E. Rudd, Nucl. Phys. {\bf B352} (1991) 489.\hb
C. Duval, G.W. Gibbons and P.A. Horvathy, Phys. Rev. {\bf D43} (1991) 3907.\hb
A.A. Tseytlin, Nucl. Phys. {\bf B390} (1993) 153; Phys. Rev. {\bf D47}
(1993) 3421.\hb
C. Duval, Z. Horvath and P.A. Horvathy, Phys. Lett. {\bf B313} (1993) 10.\hb
E.A. Bergshoeff, R. Kallosh and T. Ortin, Phys. Rev. {\bf D47} (1993) 5444.}
\lref\SANT{J. H. Horne, G.T. Horowitz and A. R. Steif,
Phys. Rev. Lett. {\bf 68} (1991) 568.}

\lref\PRE{J. Prescill, P. Schwarz, A. Shapere, S. Trivedi and F. Wilczek,
Mod. Phys Lett. {\bf A6} (1991) 2353.\hb
C. Holzhey and F. Wilczek, Nucl. Phys. {\bf B380} (1992) 447.}

\lref\HAWK{J. B. Hartle and S. W. Hawking Phys. Rev. {\bf D13} (1976) 2188.\hb
S. W. Hawking, Phys. Rev. {\bf D18} (1978) 1747.}

\lref\HAWKI{S. W. Hawking, Comm. Math. Phys. {\bf 43} (1975) 199.}

\lref\HAWKII{S. W. Hawking, Phys. Rev. {\bf D14} (1976) 2460.}

\lref\euclidean{S. Elitzur, A. Forge and E. Rabinovici,
Nucl. Phys. {\bf B359} (1991) 581. }

\lref\ITZ{C. Itzykson and J. Zuber, {\it Quantum Field Theory},
McGraw Hill (1980). }

\lref\kacrev{P. Goddard and D. Olive, Journal of Mod. Phys. {\bf A} Vol. 1,
No. 2 (1986) 303.}

\lref\BBS{F.A. Bais, P. Bouwknegt, K.S. Schoutens and M. Surridge,
Nucl. Phys. {\bf B304} (1988) 348.}

\lref\nonl{A. Polyakov, {\it Fields, Strings and Critical Phenomena}, Proc. of
Les Houses 1988, eds. E. Brezin and J. Zinn-Justin North-Holland, 1990.\hb
Al. B. Zamolodchikov, preprint ITEP 87-89. \hb
K. Schoutens, A. Sevrin and P. van Nieuwenhuizen, Proc. of the Stony Brook
Conference {\it Strings and Symmetries 1991}, World Scientific,
Singapore, 1992. \hb
J. de Boer and J. Goeree, ``The Effective Action of $W_3$ Gravity to all
\hb orders'', THU-92/33.}

\lref\HOrev{G.T. Horowitz, {\it The Dark Side of String Theory:
Black Holes and Black Strings}, Proc. of the 1992 Trieste Spring School on
String Theory and Quantum Gravity.}

\lref\HSrev{J. Harvey and A. Strominger, {\it Quantum Aspects of Black
Holes}, Proc. of the 1992 Trieste Spring School on
String Theory and Quantum Gravity.}

\lref\GM{G. Gibbons, Nucl. Phys. {\bf B207} (1982) 337.\hb
G. Gibbons and K. Maeda, Nucl. Phys. {\bf B298} (1988) 741.}

\lref\GID{S. B. Giddings, Phys. Rev. {\bf D46} (1992) 1347.}

\lref\PRErev{J. Preskill, {\it Do Black Holes Destroy Information?},
Proc. of the International Symposium on Black Holes, Membranes, Wormholes,
and Superstrings, The Woodlands, Texas, 16-18 January, 1992.}

\lref\tye{S-W. Chung and S. H. H. Tye, Phys. Rev. {\bf D47} (1993) 4546.}

\lref\eguchi{T. Eguchi, Mod. Phys. Lett. {\bf A7} (1992) 85.}

\lref\blau{M. Blau and G. Thompson, Nucl. Phys. {\bf B408} (1993) 345.}

\lref\HSBW{P. S. Howe and G. Sierra, Phys. Lett. {\bf 144B} (1984) 451.\hb
J. Bagger and E. Witten, Nucl. Phys. {\bf B222} (1983) 1.}

\lref\GSW{M. B. Green, J. H. Schwarz and E. Witten, {\it Superstring Theory},
Cambridge Univ. Press, Vols. 1 and 2, London and New York (1987).}

\lref\KAKU{M. Kaku, {\it Introduction to Superstrings}, Springer-Verlag, Berlin
and New York (1991).}

\lref\LSW{W. Lerche, A. N. Schellekens and N. P. Warner, {\it Lattices and
Strings }, Physics Reports {\bf 177}, Nos. 1 \& 2 (1989) 1, North-Holland,
Amsterdam.}

\lref\confrev{P. Ginsparg and J. L. Gardy in {\it Fields, Strings, and
Critical Phenomena}, 1988 Les Houches School, E. Brezin and J. Zinn-Justin,
eds, Elsevier Science Publ., Amsterdam (1989). \hb
J. Bagger, {\it Basic Conformal Field Theory},
Lectures given at 1988 Banff Summer Inst. on Particle and Fields,
Banff, Canada, Aug. 14-27, 1988, HUTP-89/A006, January 1989. }

\lref\CHAN{S. Chandrasekhar, {\it The Mathematical Theory of Black Holes},
Oxford University Press, 1983.}

\lref\KOULU{C. Kounnas and D. L\"ust, Phys. Lett. {\bf 289B} (1992) 56.}

\lref\PERRY{M. J. Perry and E. Teo, ``Non-singularity of the Exact two
Dimensional String Black Hole'', DAMTP-R-93-1. \hb
P. Yi, ``Nonsingular 2d Black Holes and Classical String Backgrounds'',
CALT-68-1852. }

\lref\GiKi{A. Giveon and E. Kiritsis, Nucl. Phys. {\bf B411} (1994) 487.}

\lref\kar{S. K. Kar and A. Kumar, Phys. Lett. {\bf 291B} (1992) 246.}

\lref\NW{C. Nappi and E. Witten, Phys. Rev. Lett. {\bf 71} (1993) 3751.}

\lref\HK{M. B. Halpern and E. Kiritsis,
Mod. Phys. Lett. {\bf A4} (1989) 1373.}

\lref\MOR{A.Yu. Morozov, A.M. Perelomov, A.A. Rosly, M.A. Shifman and
A.V. Turbiner, Int. J. Mod. Phys. {\bf A5} (1990) 803.}

\lref\KK{E. Kiritsis and C. Kounnas, Phys. Lett. {\bf B320} (1994) 264.}

\lref\KST{K. Sfetsos and A.A. Tseytlin,
``Antisymmetric tensor coupling and conformal
invariance in sigma models corresponding to gauged WZNW theories'',
THU-93/25, CERN-TH.6969/93, hepth/9310159,
to appear in Phys. Rev. {\bf D} (1994).}

\lref\KSTh{K. Sfetsos and A.A. Tseytlin,
``Chiral gauged WZNW models and heterotic string
backgrounds'', CERN-TH.6962/93, USC-93/HEP-S2, hepth/9308018,
to appear in Nucl. Phys. {\bf B} (1994).}

\lref\etc{K. Sfetsos, ``Gauging a non-semi-simple WZW model'',
THU-93/30, hepth/9311010, to appear in Phys. Lett. {\bf B} (1994).}

\lref\KT{C. Klimcik and A.A. Tseytlin, ``Duality invariant class of
exact string backgrounds'', CERN TH.7069, hepth/9311012.}

\lref\saletan{E.J. Saletan, J. Math. Phys. {\bf 2} (1961) 1.}
\lref\jao{D. Cangemi and R. Jackiw, Phys. Rev. Lett. {\bf 69} (1992) 233.}
\lref\jat{D. Cangemi and R. Jackiw, Ann. Phys. (NY) {\bf 225} (1993) 229.}

\lref\ORS{ D. I. Olive, E. Rabinovici and A. Schwimmer, ``A class of String
Backgrounds as a Semiclassical Limit of WZW Models'', SWA/93-94/15,
hepth/93011081.}

\lref\edc{K. Sfetsos, ``Exact String Backgrounds from WZW models based on
Non-semi-simple groups'', THU-93/31, hepth/9311093, to appear in
Int. J. Mod. Phys. {\bf A} (1994).}

\lref\grnonab{A. Giveon and M. Rocek, ``On non-abelian duality'',
ITP-SB-93-44, RI-152-93, hepth/9308154.}

\lref\wigner{E. ${\rm In\ddot on\ddot u}$ and E.P. Wigner,
Proc. Natl. Acad. Sci. U. S. {\bf 39} (1953) 510.}

\lref\TSEma{A.A. Tseytlin, ``On a Universal class of WZW-type
conformal models'', CERN-TH.7068/93, hepth/9311062.}

\lref\BCHma{J. de Boer, K. Clubock and M.B. Halpern, ``Linearized form
of the Generic Affine-Virasoro Action'', UCB-PTH-93/34, hepth/9312094.}

\lref\AABL{E. Alvarez, L.Alvarez-Gaume, J.L.F. Barbon and Y. Lozano,
``Some global aspects of Duality in String Theory'', CERN-TH.6991/93,
FTUAM.93/28, hepth/9309039.}

\lref\noumo{N. Mohammedi, ``On Bosonic and Superconformal Current Algebra
for Non-semi-simple Groups'', BONN-HE-93-51, hepth/9312182.}

\lref\figsonia{J.M. Figueroa-O'Farrill and S. Stanciu,
``Nonsemisimple Sugawara constructions'', QMW-PH-94-2, hepth/9402035.}


\hfill {THU-94/01}
\vskip -.3 true cm
\rightline{February 1994}
\vskip -.3 true cm
\rightline {hep-th/9402031}

\bs\bs\bs

\centerline  {\bf GAUGED WZW MODELS AND    }
\centerline  { \bf NON-ABELIAN DUALITY }

\vskip 1.00 true cm

\centerline  {  {\bf Konstadinos Sfetsos}{\footnote{$^*$}
 {e-mail address: sfetsos@fys.ruu.nl }}                                     }

\bigskip

\centerline {Institute for Theoretical Physics }
\centerline {Utrecht University}
\centerline {Princetonplein 5, TA 3508}
\centerline{ The Netherlands }


\vskip 1.30 true cm

\centerline{ABSTRACT}

\vskip .3 true cm

We consider WZW models based on the non-semi-simple algebras that they
were recently constructed as contractions of corresponding algebras for
semi-simple groups. We give the explicit expression for the action of these
models, as well as for a generalization of them, and discuss their general
properties.
Furthermore we consider gauged WZW models based on these non-semi-simple
algebras and we show that there are equivalent to non-abelian duality
transformations on WZW actions.
We also show that a general non-abelian duality transformation
can be thought of as a limiting case of the non-abelian quotient theory
of the direct
product of the original action and the WZW action for the symmetry gauge
group $H$. In this action there is no Lagrange multiplier term
that constrains the gauge field strength to vanish. A particular result is
that the gauged WZW action for the coset $(G_k \otimes H_l)/H_{k+l}$ is
equivalent, in the limit $l\to \infty$, to the dualized WZW action for
$G_k$ with respect to the subgroup $H$.

\vfill\eject


\newsec{ Introduction }

One of the most striking symmetries in String theory is that of duality.
The simplest example of a theory with this symmetry is that of a single
boson compactified on a circle of radious $R$ \rovr. From the mathematical
point of view the partition function of the theory is invariant under
the duality transformation $R\to 1/R$ (in Planck units) for all genera
(a redefinition of the constant dilaton is also necessary to insure the
invariance of the string coupling constant)
and from a more physical point of view invariance under duality
implies that Physics at small scales is indistinguishable from Physics at large
scales and that a smaller distance beyond which probing Physics does not make
sense should exist \rovr\GV\nssw.
Duality is not a property of point-like objects.
Moreover it is exclusively stringy in the sense that it
is not also a property of higher than one dimensional
extended objects ($p$-branes) \vafa.

This simple case was generalized to arbitrary toroidal compactifications
\narain\nssw\grvm\
and subsequently it was realized, at the non-linear $\s$-model level,
that duality was a symmetry of all string vacua with one
\BUSCHER\ or more abelian isometries \RV\GR\GIN.
In the case of $d$ abelian isometries the duality transformations enlarged to
$O(d,d,R)$ ones relate apparently different curved backgrounds in
String theory and can be used to generate new solutions from known ones
\Dualone\Dualtwo.
Also duality plays an important role in discussing Cosmology in
the context of String theory \bv\gsvy\tv.
It has been argued \RV\ that in the case of $d$ abelian isometries with
compact orbits the duality group $O(d,d,Z)$ \GR\
interpolates between different
backgrounds that are manifestations of the same Conformal Field Theory (CFT)
(for the non-compact $SL(2,\IR)/\IR$ case see \IBCS\slt\KIR\DVV\GiKi).
An important feature is that the dual of such background has also the same
number of abelian isometries and that its dual is the original model.
For the relation between abelian duality transformations and marginal
perturbations of String backgrounds see \dualmargi\ and for review articles
see \KIRd\gpr.

There is a second kind of duality transformations which is much less
understood, where the isometries with respect to which the dualization is done
form a non-abelian group \nadual\ (for earlier work see \duearl).
This has very important consequences.
First of all generically the dual model has much less symmetry that the
original one and the isometry group usually disappears or gets smaller
\nadual\grnonab\AABL\GRV\
(in fact the local original symmetry seems to manifest
itself in a non-local way in the dual model \grnonab).
Moreover, it has been argued that the non-abelian duality
transformations interpolate between solutions not of the same CFT but of
different ones possibly related by orbifold construction
\nadual\grnonab\ (for an extensive treatment of global issues on duality
transformations see \AABL).
The analog of the duality group of $O(d,d,Z)$ for abelian duality is not
known and in fact because the initial isometry group is not preserved, one
does not know how to find the `inverse' transformation.

A common characteristic of all duality transformations is that they can be
formulated in a way that the action is that of the original theory written in a
gauge invariant way by using gauge fields and a Lagrange multiplier term
that constrains, upon integration in the path integral,
the gauge field
strength to vanish \duearl, thus giving (after gauge fixing) the original model
\RV\nadual.
The dual theory is obtained by integrating instead over the
gauge fields. At this point the procedure resembles the one that is being
extensively followed in the case of gauged WZW models \GWZW,
when the original theory is a WZW model \Wwzw,
or more generally in the case of non-abelian quotient models where a target
spacetime symmetry is being gauged.
For instance, similarly to what we have already mentioned for the case
of non-abelian duality transformations, in the case of gauged WZW models the
original global symmetry that is being gauged also disappears, i.e. it
is being gauge fixed.
Therefore one may wonder whether or not this apparent similarity can be made
more precise. It will be explained that such a relationship is found in the
context of contraction of certain non-abelian quotient models and on
gauged WZW models based on a particular class of non-semi-simple groups.

Recently WZW models based on non-semi-simple groups have been constructed
\NW\ORS\edc. It was shown in \ORS\ that the symmetry current algebra for
these models can be obtained by a contraction procedure on the current algebra
of the WZW model for the direct product group $G\otimes H$.
A feature all of these models have in common is that their central charge is
an integer. A bosonization and computation of
the spectrum of the first and the simplest of these theories \NW\ with
$G=SO(3)$ and $H=SO(2)$ and denoted by $E^c_2$ can be found in \KK.
Gauged WZW models were also constructed
by gauging various anomaly free subgroups of $E^c_2$ \KK\etc.
In particular the $3D$ model in \etc\ was shown to correspond to a correlated
limit of the charged black string background
$SL(2,\IR) \otimes \IR/\IR$ of \HOHO\SFET\
both in the semiclassical limit as well as when all $\a'$-corrections are
included. Moreover, although curvature singularities are still present it was
shown that they can be removed via abelian duality transformations
that map this background to a flat spacetime with constant antisymmetric tensor
and dilaton fields (for the neutral black string, that is directly related to
the $2D$ black hole \WIT, a similar conclusion holds).
One of the aims of this paper is to generalize this us
much as possible to the general models of \ORS.

The organization of the paper and some of the main results are:
In section 2 in order to set up our notation and for completeness
first we review the current algebra construction of \ORS\
for WZW models based on a class of non-semi-simple algebras as a particular
correlated limit of the current algebra for the direct product $G\otimes H$.
Next we derive an explicit form of the corresponding WZW action that reveals
all the essential properties of these models, such as the existence of $dim(H)$
null Killing vectors, by generalizing the work of \edc. We also explicitly
show how this action can be obtained from a correlated limit on
the WZW action for $G\otimes H$.
In section 3 we formulate gauged WZW models by gauging an anomaly free
subgroup. Before the contraction this corresponds to the gauged WZW models
for the cosets $(G\otimes H)/H$ and we show the explicit correspondence.
Unlike the original WZW models who have integer central charges the gauged
models have in general rational central charges.
Section 4 contains two main results.
First, general non-abelian duality transformations are shown to correspond to
limiting cases of direct product models,
where a target spacetime symmetry is being gauged. In these models before the
limit is taken the gauge field strength is {\it not} constrained to vanish.
As a particular case non-abelian transformations on the WZW model for a group
$G$ with respect to the vectorial action of a subgroup $H$ can be thought of
as a limit of the gauged WZW model for the coset $(G\otimes H)/H$.
The second result of this section is that the gauged WZW models of section 3
are equivalent to non-abelian duality transformations on the WZW model for
the direct product $H\otimes U(1)^{dim(G/H)}$.
Our formulation of non-abelian duality
makes possible to compute the $\a'$-corrections to the semiclassical
expressions for the $\s$-model background fields,
by making contact with known results from the coset models.
In all cases if $H=G$ all $\a'$-corrections vanish even though the
corresponding $\s$-models are non-trivial.
We end the main part of the paper with concluding remarks and discussion of
our results in section 5. In Appendix A we extend the
results of section 3 to the case of axial gauging and we show that the
resulting curved backgrounds can be obtained from the flat one with constant
antisymmetric tensor and dilaton fields. In Appendix B we extend the
construction of \ORS\ to cover more general cases. In Appendix C we present
a CFT description for the generalization of the plane wave
model of \NW\ in higher dimensions.

\newsec{WZW models based on non-semi-simple groups}

In this section we will construct WZW models based on {\it non-semi-simple}
groups. This will be done by first reviewing the work of \ORS\ on the
construction of the current algebra for such models via a contraction of the
current algebra for WZW models based on {\it semi-simple} groups.
Then we will give explicit expressions for the action of the resulting WZW
$\s$-models by following \edc. Many other formulae derived in this section
will be useful in subsequent ones.


Let us consider the WZW model for the direct product group $G\otimes H$,
where $G$ and $H$ are groups
(throughout this section they are taken to be compact ones)
and where $G$ should contain a subgroup isomorphic
to $H$. The holomorphic currents associated with the current algebra symmetry
of the corresponding WZW model are $\hat g=\{u_i, R_{\a}\}$,
$\hat h=\{v_i\}$, where $i=i,2,\dots, dim(H)$ and $\a=1,2,\dots, dim(G/H)$.
They obey the Operator Product Expansions (OPE's)\foot{Throughout this paper
OPE's that are not written down explicitly are assumed to have
only regular terms.
Also we will not explicitly mention the various Lie algebras
since one can easily read them from the associated OPE's.
Moreover, we will use the same symbols
for Lie algebra and current algebra generators. }
\eqn\opea{\eqalign{&u_i u_j \sim {i f_{ij}{}^k u_k\ov z-w}+{k_G\ \eta_{ij}
\ov (z-w)^2}\ ,\qq v_i v_j \sim {i f_{ij}{}^k v_k\ov z-w}+{k_H\ \eta_{ij}
\ov (z-w)^2}  \cr
&R_{\a}R_{\b} \sim {i M_{\a\b}{}^iu_i + i S_{\a\b}{}^{\g} R_{\g}\ov z-w}
+{k_G\ \eta_{\a\b}\ov (z-w)^2}\ ,
\qq u_iR_{\a} \sim {i M_{i\a}{}^{\b} R_{\b}\ov z-w}\ ,\cr}}
where $f_{ij}{}^k$, $M_{\a\b}{}^i$ and $S_{\a\b}{}^{\g}$ are structure
constants of the corresponding Lie algebras one can use to compute the
Killing metrics $\eta_{ij}$, $\eta_{\a\b}$. The levels $k_G$ and $k_H$
are assumed to be positive integers.
The energy momentum tensor and the central charge of the
corresponding Virasoro algebra are
\eqn\vira{T={:u^2 +R^2:\ov 2(k_G+g_G)} + {:v^2: \ov 2(k_H+g_H)}\ ,\qq
c={k_G\ dim(G)\ov k_G + g_G} + {k_H\ dim(H) \ov k_H + g_H}\ ,}
where $g_G$, $g_H$ are the dual Coxeter numbers for $G$ and $H$ and where the
regularization prescription of \BBS\ is assumed in writing current bilinears.
As in \ORS\ we let
\eqn\conta{\eqalign{
&T_i=u_i+v_i\ ,\qq F_i=\e (u_i-v_i)\ ,\qq P_{\a}= \sqrt{2\e}\ R_{\a}\cr
&k_G=\ha (\b+\a/\e)\ ,\qq k_H=\ha (\b-\a/\e)\ ,\cr}}
and rewrite \opea\ in the basis $\{J_A\}=\{P_{\a},T_i,F_i\}$.
In the limit $\e\to 0$ we discover a new current algebra
{\it not } equivalent to \opea, because the transformation \conta\ is not
invertible in that limit. Let us also note that $\b\in \IR^+$ and $\a\in \IR$.
The OPE's of the new current algebra one obtains are
\eqn\opeac{\eqalign{
&T_iT_j\sim {if_{ij}{}^k T_k\ov z-w} + {\b\ \eta_{ij} \ov (z-w)^2}\ ,
\qq T_i P_{\a}\sim {iM_{i\a}{}^{\b}P_{\b}\ov z-w} \cr
&T_i F_j\sim {if_{ij}{}^k F_k\ov z-w} + {\a\ \eta_{ij}\ov (z-w)^2}\ ,
\qq P_{\a}P_{\b}\sim {i M_{\a\b}{}^i F_i\ov z-w} + {\a\ \eta_{\a\b}\ov (z-w)^2}
\ .\cr} }
We will denote this current algebra by ${\hat g}^c_h$ and the corresponding
Lie algebra by $g^c_h$ (for the group $\ghc$ will be used).
The holomorphic stress tensor and the central charge of the
corresponding Virasoro algebra obtained from \vira, using
\conta\ in the $\e\to 0$ limit, read
\eqn\virac{T={:P^2 +2 FT:\ov 2\a} - {\b+g_G + g_H\ov 2 \a^2}\ :F^2:\ , \qq
c=dim(G)+dim(H)\ .}
Of course with respect to the above energy momentum tensor all currents are
primary fields of conformal dimension one.
The OPE's in \opeac\ define a quadratic form
\eqn\quada{\Om_{AB}=\bordermatrix{ & P_{\b} & T_j & F_j \cr
P_{\a} & \a/\b\ \eta_{\a\b} & 0& 0\cr
T_i & 0 &  \eta_{ij} &\a/\b\ \eta_{ij}\cr
F_i & 0 & \a/\b\ \eta_{ij} & 0 \cr }\ ,}
which is symmetric, i.e. $\Omega_{AB}=\Omega_{BA}$,
a group invariant, i.e. $f^D_{AB}\Omega_{CD} + f^D_{AC}\Omega_{BD} = 0$
and the inverse matrix
\eqn\iquada{\Om^{AB}=\bordermatrix{ & P_{\b} & T_j & F_j \cr
P_{\a} & \eta^{\a\b} & 0& 0\cr
T_i & 0 & 0 & \eta^{ij} \cr
F_i & 0 & \eta^{ij}  & -\b/\a\ \eta^{ij}\cr } \b/\a\ ,}
obeying $\Omega^{AB}\Omega_{BC}=\eta^A_C$ exists.
The above properties of the quadratic form \quada\ are a consequence of the
fact that the current algebra \opeac\ is a contraction of \opea\ for
which the Killing metric $\eta_{AB}$, sharing all of these properties,
is taking as the quadratic form.\foot{For the case of $G=SO(3)$ and $H=SO(2)$
the Lie algebra that \opeac\ defines appeared before in the context of
contraction of
Lie groups \saletan\ and in studies of $(1+1)$-dimensional gravity \jao.
For the same case the quadratic form \quada\ had appeared in \jat.}
Nevertheless one may explicitly verify them.

A form for the WZW action whose symmetry algebra is \opeac\ was given for the
general model in \ORS. However, it is not very explicit
(in particular it involves two Wess-Zumino terms) and extracting general
conclusions from it is rather difficult. Here we will
follow the method, applied explicitly for the case of $E_d^c$
(in our notation $E^c_d = SO(d+1)^c_{so(d)}$) models in \edc,
which involves an explicit parametrization of the group element $g\in \ghc$.
The latter can be generally
parametrized as (summation over repeated indices is implied)
\eqn\param{g=e^{i a\cdot P} e^{i v\cdot F} h_x\ ,\qq a\cdot P=a^{\a}P_{\a}\ ,
\qq v\cdot F=v^i F_i\ ,}
where the group element $h_x\in H$ parametrizes
$dim(H)$ parameters $x^{\m}$ and the $a^{\a}$'s and $v^i$'s the remaining ones.
It will be useful at this point to further establish
our notation by introducing some useful matrices
\eqn\nota{\eqalign{
&C_{ij}(h_x)=\Tr(T_i h_x T_j h\inv_x)=\b/\a\ \Tr(F_i h_x T_j h\inv_x)\ ,
\qq C_{ik} C_j{}^k=\eta_{ij} \cr
&L^i_{\m}=-i\ \Tr(T^i h\inv_x \partial_{\m} h_x)\ ,
\qq R^i_{\m}=-i\ \Tr(T^i \partial_{\m}h_x h\inv_x )=C^i{}_j L^j_{\m}\cr
&M_{ij}=C_{ij}-\eta_{ij}\ ,\qq m_{i\a}=M_{i\b\a}a^{\b}\ ,
\qq n_{ij}=f_{ij}{}^k v_k\ .\cr} }
We will also denote by $L^{\m}_i$, $R^{\m}_i$ the inverses of the matrices
$L^i_{\m}$, $R^i_{\m}$ respectively, and by $m^t$ the transpose matrix of $m$.
Using the Duhamel's formula
\eqn\form{de^H=\int_0^1 ds\ e^{s H} dH e^{(1-s) H} }
one can compute the left-invariant Maurer-Cartan form
\eqn\lj{g\inv dg= i(da\cdot P' -\ha da^{\a} m_{\a}{}^iF'_i)
+i dv\cdot F' + iR_{\m}^i dx^{\m} T'_i\ ,}
where the generators $J'_A=h_x\inv J_A h_x$ satisfy the same commutation
relations as the $J_A$'s. Similarly we compute the right-invariant
Maurer-Cartan form
\eqn\rj{dgg\inv=
i(da\cdot P +\ha da^{\a} m_{\a}{}^i F_i)
+i R^i_{\m} dx^{\m} e^{i\a\cdot P} e^{iv\cdot F} T_i
e^{-iv\cdot F} e^{-i\a\cdot P} +i  dv\cdot F\ .}
To obtain an explicit expression we note that $dgg\inv=-gdg\inv$ and then we
make use of
\lj\ with ($a_i\to -a_i$, $ v_i\to -v_i$, $h_x\to h_x\inv$) and
($P_i\to P_i'$, $F_i\to F'_i$) with the additional
contribution of the terms $a\cdot dP'$ and $v\cdot dF'$ (these terms contribute
when derivatives with respect to the parameters $x^{\m}$ of $h_x$ are taken).
The resulting expression is
\eqn\dggi{\eqalign{dgg\inv=&ida^{\a}(P_{\a} +\ha m_{\a}{}^iF_i)
+ idv\cdot F  + iR^i_{\m} dx^{\m} \bl(T_i + m_i{}^{\a} P_{\a}
+(\ha m_i{}^\a m_\a{}^j - n_i{}^j )F_j \br)\ .\cr} }
Using the previous expression for the right-invariant Maurer-Cartan form
we compute the following matrix defined as $dgg\inv=idX^M E_M{}^A J_A$,
where $X^M=\{x^{\m}, a^{\a}, v^i\}$
\eqn\ema{E_M{}^A=\bordermatrix{ & P_{\b} & T_i & F_i \cr
x^{\m} & R^k_{\m} m_k{}^{\b} & R^i_{\m}
& R^k_{\m}(\ha m_k{}^{\g}m_{\g}{}^i-n_k{}^i)\cr
a^{\a} &\d_{\a}{}^{\b} & 0& \ha m_{\a}{}^i \cr
v^j & 0 & 0 & \d_j{}^i \cr } }
and its inverse
\eqn\eam{E_A{}^M=\bordermatrix{ & x^{\m} & a^{\a} & v^j \cr
P_{\b} & 0 & \d_{\b}{}^{\a} & -\ha m_{\b}{}^j \cr
T_i & R^{\m}_i & -m_i{}^{\a} & n_i{}^j \cr
F_i  & 0 & 0 & \d_i{}^j \cr }\ .}
The zero modes of the holomorphic currents $J_A$ can be constructed as first
order differential operators acting on the group parameter space of $G$ using
$J_A=i E_A{}^M \partial_M$. Their explicit form is
\eqn\oper{\eqalign{
&P_{\a}=i \partial_{a^{\a}} - {i\ov 2} m_{\a}{}^j \partial_{v^j}\cr
& T_i= i R^{\m}_i \partial_{x^{\m}} - i m_i{}^{\b} \partial_{a^{\b}}
+ i n_i{}^j \partial_{v^j} \cr
&F_i= i \partial_{v^i} \ .\cr} }
Using the fact that $i R^{\m}_i \partial_{x^{\m}}$ separately obey $\cL(H)$,
i.e. the Lie algebra of the group $H$,
one may explicitly verify, with the aid of the Jacobi identities that the
various structure constants obey, that the operators \oper\ indeed
generate the Lie algebra $g^c_h$.
The WZW action whose current algebra symmetry is \opeac\ is defined as
\eqn\wzw{S(g)={\b\ov 2\pi}\int_\Sigma d^2z\ T(\del g\inv \bd g \Omega)
+{\b\ov 6\pi}\int_B \Tr (g\inv dg \wedge g\inv dg \wedge g\inv dg \Omega )\ ,}
where $\Omega_{AB}=Tr(T_A T_B \Omega)$ and $\Sigma=\del B$. Using a procedure
analogous to the one in \edc\ or by using the Polyakov-Wiegman
identity \POLWIG\
(the latter is valid in our case because of the properties of the quadratic
form $\Om_{AB}$) and \dggi\ we evaluate
\eqn\wzwc{S(g)=\b\ I_0(h_x) +{\a\ov 2\pi}\int_{\Sigma} d^2z\ [\
\del a\cdot \bd a + (2 \del v_i
+ m_{i\a}\del a^{\a}) R^i_{\m} \bd x^{\m}\ ]\ ,}
where $I_0(h_x)$ is the WZW action for the group element
$h_x \in H$. It is clear that the above action corresponds to
string backgrounds, in $dim(G)+dim(H)$ space-time dimensions,
with $dim(H)$ null
Killing vectors corresponding to the coordinates $v^i$, thus generalizing the
similar statement made previously for the cases of the WZW models for
$E_2^c$ \NW\ and $E_d^c$ \edc.
The signature of the spacetime has for $\a>0$ ($\a<0$)
$dim(G)$ positive (negative) entries and $dim(H)$ negative (positive)
ones as it can be seen from the eigenvalues of the quadratic form \quada.
To obtain models with one timelike coordinate we should take $\a>0$ and
consider only one dimensional subgroups $H$.
If among the set of
generators $\{R_{\a}\}$ there is an invariant subset with corresponding
structure constants $M_{\a\b}{}^i=0$,
i.e. if the subgroup $H$ is not the maximum one,
then after the contraction \conta\ they correspond to commuting generators
or to free decoupled fields in the WZW action \wzwc\ and therefore we can
safely ignore them (for an example see Appendix C).
The correct
measure in the path integral for the action \wzwc\ is the Haar measure for
the group $H$ times the flat measure for the sets of coordinates
$\{v^i\}$ and $\{a^{\a}\}$,
as one can see by explicitly computing the square root of the determinant of
the metric corresponding to \wzwc. Even though the constant $\b$ as defined in
\conta\ can be any positive real number it should be chosen to be
a positive integer in order to have a well defined path integral for
\wzwc\ \Wwzw.
It is a also a quite straightforward
computation to verify,
using the stress tensor \virac\ and the Lie algebra differential operators
\oper,
that the same metric (up to a shift $\b\to \b + g_G +g_H$),
and a constant dilaton arises
using the operator algebraic method of \DVV\BSexa.\foot{The forementioned
shift in the value of the constant $\b$ is a direct consequence of the
regularization prescription used in \virac\ and obviously does not affect
the vanishing of the $\b$-functions in conformal perturbation
theory (see for instance \CALLAN).
Also notice that the fact that
the central charge in \virac\ does not depend at all on $\b$ means that
no matter what regularization scheme we use
all loop contributions to it should vanish.}

Being a WZW action \wzwc\ has a number of `obvious'
global symmetries corresponding to the transformations
\eqn\glo{ h_x\to h_x \Lambda\ ,\qq \{ h_x\to S h_x,\ v\to S v S\inv,\
a\to S a S\inv \}\ ,\qq  v\to v+N\ ,}
where $\Lambda$, $S$ are constant group elements of $H$ and $N$ a constant
matrix in $\cL(H)$. The first two transformations in \glo\
represent `left' and `right' glogal transformations of the group element in
\param\ and the third one is due to the fact that there exist Killing vectors
along the directions $v^i$. One might use these symmetries to generate new
solutions via duality transformations.

We could have obtained \opeac-\quada\ with
a different logic (this was effectively used for the cases of $E^c_2$
and $E^c_d$ in \NW\edc).
If one starts with the Lie algebra of the group $G$
performs a ${\rm In\ddot on\ddot u}$-Wigner contraction \wigner,
i.e. $R_{\a}\to {1\ov \e} R_{\a}$, then one discovers that the WZW action
one writes down using the Killing metric cannot be anything else
but that of the WZW model for the subgroup $H$ itself (this is because the
Killing metric is degenerate due to the non-semi-simplicity of the
corresponding contracted algebra).
If one insists that all the currents are primary fields of
conformal weight one then the minimal
resolution to the problem
(for a different and more involved one see Appendix B)
is to introduce extra generators
$\{F_i\}$ with the OPE's given in a unique way by \opeac\ which correspond to
the non-degenerate quadratic form \quada, thus making it possible to write
the corresponding WZW action. However, this approach is not as nice as
the one in \ORS\ since it does not make contact with already existing models.
If the condition that all the currents must be primaries is relaxed,
then we can still have a consistent current algebra given by \opeac\ with the
$F_i$'s taken formally to zero.
Then one can show that there exist a stress tensor and a corresponding central
charge given by
\eqn\virstr{T={:P^2:\ov 2\a} + {:T^2:\ov 2(\b +g_H)}\ ,\qq
c=dim(G) + {g_G-2 g_H\ov \b + g_H}\ dim(H)\ .}
However, only the $T_i$'s and not the $P_{\a}$'s are
primary fields with conformal dimension one with respect to that stress tensor.
Because of this there can be no WZW action with the symmetry algebra we just
described. A similar conclusion arises from the fact that the quadratic form
for this theory (given by the relevant entries in \quada) even though it is
invertible and symmetric it is not a group invariant and therefore
the Wess-Zumino term, necessary for any WZW model, cannot be defined.
Nevertheless, because \virstr\ is a solution of the Master equation \HK\MOR\
(a new one, to the best of our knowledge)
it is conceivable that an explicit form for the corresponding
action using the results of \TSEma\BCHma, can also be found.

One may wonder whether or not it is possible to obtain the action
\wzwc\ through a limiting procedure taken directly at the level of the
WZW action for the direct product group $G\otimes H$. After all this is how
one obtains the OPE's in \opeac\ in a natural way.
Let us start with the action for the $G\otimes H$
(with a general parametrization for the group elements)
\eqn\unc{ S=k_G\ I_0(t h_{x'}) + k_H\ I_0(h_x)\ ,\qq t=e^{i a\cdot R}\ ,}
where $t$ belongs in the left coset, i.e. $t\in \bl(G/H\br)_L$, and $h_{x'}$
and $h_x$ contain the remaining subgroup variables.
Next we expand the subgroup element $h_{x'}$ near the corresponding element
$h_x$ as $h_{x'}=(I+2 i \e\ v\cdot u) h_x$, where by the $v^i$'s we
collectively call the result of various, generically more complicated,
shifts and rescalings of the original parameters $x^{\m}$ and\foot{If we write
$h_{x'}=e^{ix'\cdot u}$ (and similarly for $h_x$) and let
$x'_i=x_i+2 i\e \phi_i$ we can easily show using \form\ and \nota\ that
in the limit $\e\to 0$
$$v^i= \phi^j \int_0^1 ds\ C^i{}_j(h_{sx})\ , \qq h_{sx}=e^{i s x\cdot u}\ ,$$
where $C_{ij}$ is the matrix defined in \nota. Obviously only for abelian
subgroups $v^i=\phi^i$.}
also we scale the parameters $a^{\a}\to \sqrt{2\e}\ a^{\a}$.
Then by writing $t h_{x'}\equiv t_{\e} h_x$ where the matrix $t_{\e}$ and its
inverse are
\eqn\te{\eqalign{&t_{\e}= I + i\sqrt{2\e}\ a\cdot R
+ \e\ [\ 2 i v\cdot u -(a\cdot R)^2\ ] + O(\e^{3/2})\cr
& t_{\e}\inv = I - i\sqrt{2\e}\ a\cdot R
- \e\ [\ 2 i v\cdot u +(a\cdot R)^2\ ]  +O(\e^{3/2})\ ,\cr } }
one can compute, in powers of $\e$, the corresponding left and right-invariant
Maurer-Cartan forms and the WZW action for the group element $t_{\e}$
\eqn\cartan{\eqalign{ & t_{\e}\inv dt_{\e}= i\sqrt{2\e}\ da\cdot R +
i \e\ [\ 2 dv\cdot u  - da^{\a} m_{\a}{}^i u_i
+ \a^{\a} da^{\b} S_{\a\b}{}^{\g} R_{\g}\ ] + O(\e^2) \cr
& dt_{\e} t_{\e}\inv = i \sqrt{2\e}\ d a\cdot R + i \e\ [\ 2 d v\cdot u
+ d a^{\a} m_{\a}{}^i u_i
- \a^{\a} d a^{\b} S_{\a\b}{}^{\g} R_{\g}\ ] + O(\e^2) \cr
& I_0(t_{\e}) = {\e\ov \pi} \int_{\Sigma} d^2 z\ \del a\cdot \bd a
+ O(\e^{3/2})\ .\cr } }
Also the formulae
\eqn\car{\eqalign {
&k_G\  \Tr[\ u_i\ (th_{x'})\inv \del(t h_{x'})\ ]
+ k_H\ \Tr( v_i\ h_x\inv \del h_x) = i \b L^i_{\m} \del x^{\m}
+ i\a (\del v_j +\ha m_{j\a} \del a^{\a})\ C^{ji}\cr
& k_G\ \Tr[\ u_i \bd (th_{x'}) (t h_{x'})\inv \ ]
+ k_H\ \Tr(v_i\ \bd h_x h_x\inv )=
i\ [\ \b I + \a (n +\ha m m^t)\ ]_{ij} R^j_{\m} \bd x^{\m}\cr
&+ i\a\ (\bd v_i -\ha m_{i\a} \bd a^{\a} )\ ,\cr } }
valid in the limit $\e\to 0$, will prove useful in subsequent sections.
Using the above expansion formulae, the Polyakov-Wiegman identity, as well as
the redefinitions of $k_G$ and $k_H$ in \conta\ one can see that in the limit
$\e\to 0$ the action \unc\ reduces to that in \wzwc.

\newsec{Coset models $\ghc/H$ }

In this section we use the results of the previous one to construct
gauged WZW models based on non-semi-simple groups. In particular we gauge
the subgroup generated by the subset of generators $\{T_i\}$.
We explicitly show how all the relevant formulae at the algebraic as well as at
the action level can be obtained as limiting cases of the corresponding ones
for the $(G\otimes H)/H$ gauged WZW models.

Let us consider the gauged WZW model one obtains by gauging the diagonal
subgroup, generated by $\{u_i +v_i\}$, of the direct product group
$G\otimes H$ (the axial gauging case is considered in Appendix A).
For the corresponding CFT coset model $(G\otimes H)/H$ the energy momentum
tensor and central charge of the Virasoro algebra are given by
\eqn\coset{\eqalign{&T={:u^2 + R^2:\ov 2(k_G+g_G)} + {:v^2:\ov 2(k_H+g_H)}
-{:(u+v)^2:\ov 2(k_G +k_H + g_H)} \cr
&c={k_G\ dim(G) \ov k_G + g_G}+{k_H\ dim(H) \ov k_H + g_H}
-  {(k_G +k_H)\ dim(H)\ov k_G +k_H +g_H}\ .\cr} }
Then we use \conta\ and take the limit $\e\to 0$. The energy momentum tensor
for the resulting coset theory, denoted by $\ghc/H$, reads
\eqn\cosetc{\eqalign{&T={:P^2 +2 FT:\ov 2\a} - {\b+g_G + g_H\ov 2 \a^2}\ :F^2:
- {:T^2:\ov 2(\b + g_H)} \cr
& c= dim(G) + {g_H\ dim(H)\ov \b +g_H}\ .\cr } }
Of course one may obtain the same result by directly considering the gauging
of the subgroup generated by $\{T_i\}$, of the non-semi-simple group $\ghc$.
We also see that the central charge for the gauged WZW $\ghc/H$ model
is no longer an integer as it was in the case of the WZW model for $\ghc$,
unless the subgroup $H$ is an abelian one.

To obtain an explicit form for the gauged WZW action we start with \GWZW
\eqn\gwzw{S=\b\ [\ I_0(h\inv g \bh) - I_0(h\inv \bh) \ ]\ ,}
where $h$, $\bh$ are group elements in the subgroup $H$ generated by $\{T_i\}$
and the group element $g\in \ghc$ is parametrized as in \param.
Writing explicitly the above action with the aid of \wzwc\ we obtain
\eqn\gwzwc{S=\b I(h_x,A) + {\a\ov 2\pi}\int_{\Sigma} d^2z [
Da\cdot {\bar D}a + i (Da)^{\a} m_{\a}{}^i ({\bar D} h_x h_x\inv)_i
-2i \Tr( Dv {\bar D}h_x h_x\inv - v F_{z\z}) ]\ ,}
with $I(h_x,A)$ being the usual gauged WZW action for $H/H$
\eqn\usual{I(h_x,A)=I_0(h_x)
+{1\over \pi } \int_{\Sigma} d^2 z\ \Tr \bigl( A\,\bd h_x h_x\inv -
\bar A h_x\inv\del h_x + h_x\inv A h_x \bar A  - A \A \bigr)\ ,}
and where the gauge fields $A,\A$ take values in $\cL(H)$, and
the corresponding field strength $F_{z\z}$ and the
covariant derivatives are defined as
\eqn\orismoi{A=\del h h\inv\ ,\quad \A=\bd \bh \bh\inv\ ,
\quad F_{z\z}=\del \A -\bd A -[A,\A]\ ,\quad
D=\del -[A,\cdot\ ]\ ,\quad {\bar D}=\bd -[\A,\cdot\ ]\ .}
Naively one might have expected that \gwzw\ would have been given just by
the sum of the gauged WZW for the coset $H/H$, i.e. $I_0(h_x,A)$,
and the covariantized terms one obtains by simply replacing ordinary
derivatives by
covariant ones in the terms involving $a^{\a}$ and $v^i$ in \wzwc.
In fact, such an action is gauge invariant by itself.
However, only by the inclusion of the term involving the field strength
$F_{z\z}$ one is able to rewrite it as a sum of two {\it independent}
WZW actions, as in \gwzw, that guaranties conformal invariance.
The action \gwzwc\ is invariant under the infinitesimal gauge transformations
\eqn\gtr{\d h_x=[h_x,i\e]\ ,\quad \d a^{\a}=-m^{\a}{}_ j \e^j\ ,\quad
\d v^i=  n^i{}_j \e^j\ ,\quad \d A=-iD\e\ ,\quad \d {\bar A} =-i {\bar D}\e\ ,}
where $\e=\e^i T_i$.
To obtain the $\s$-model one integrates over the gauge fields.
A straightforward computation gives\foot{Throughout this paper whenever an
action or a dilaton
is denoted with a tilded symbol will contain the rescaled
$v^i\to \b/\a\ v^i$ and $\a^{\a}\to \sqrt{\b/\a}\ a^{\a}$.}
\eqn\smod{\eqalign{\tilde{S}_{\s}=\b\ \bigl \{\ I_0(h_x) + & {1\ov 2\pi}
\int_{\Sigma} d^2 z\ [\ \del a \cdot \bd a
+ (2 \del v_i + m_{i \a} \del a^{\a}) R^i_{\m} \bd x^{\m}  \cr
&-2 \bl (L^i_{\m} \del x^{\m} +(\del v_k +\ha m_{k \a} \del a^{\a}) C^{ki}\br )
\bl ( M +(n +\ha mm^t)C \br )\inv_{ij} \cr
& \br ((I + n +\ha m m^t)^j{}_l R^l_{\n} \bd x^{\n}
+\bd v^j -\ha m^j{}_{\b} \bd a^{\b} \br )\ ]\ \bigr \}\ .\cr } }
There is also a dilaton field induced from the finite part of the determinant
one obtains by integrating out the gauge fields in \gwzwc
\eqn\dila{\tilde{\Phi}=\ln \det \bl (M + (n +\ha mm^t)C\br ) + \Phi_0\ .}
The above forms for the action and the dilaton can be simplified considerably
by using the properties of the matrix $C_{ij}$ in \nota.
The final expressions are
\eqn\smods{\eqalign{&\tilde{S}_{\s}=  \b\ \bigl \{ - I_0(h_x)
+  {1\ov 2\pi} \int_{\Sigma} d^2 z\ [\ \del \a \cdot \bd a \cr
&+2 (R^i_{\m} \del x^{\m} + \del v^i + \ha m^i{}_{\a} \del a^{\a})
(M^t - n - \ha mm^t)\inv_{ij} (L^j_{\n} \bd x^{\n}
+ \bd v^j - \ha m^j{}_{\b} \bd a^{\b})\ ] \bigr \}\ \cr }}
and
\eqn\dilas{\tilde{\Phi}= \ln \det \bl (M^t - n - \ha m m^t \br ) + \Phi_0\ .}
The signature of the spacetime described by \smods\ will have
(if $G$ and $H$ are compact groups) for $\a>0$
$dim(G/H)$ positive and $dim(H)$ negative entries and for $\a<0$ $dim(G)$
negative and no positive ones (analytically continuing $\b$ to
negative values renders a positive signature spacetime). Obviously for models
with one timelike coordinate the subgroup $H$ has to be one dimensional
and $\a,\b > 0$. If we allow the use of non-compact groups more
possibilities exist. We find that one timelike coordinate is also possible if:
1) $\b>0$, $H=G$ and $G$ is a non-compact group with one timelike generator,
i.e. $G=SO(1,d)_{-\b}$,
2) $\a>0$, $\b<0$, $H$ a compact group and $G$ a non-compact one such that
the coset $G/H$ gives rise to a spacetime with one timelike coordinate.
All such models have been classified in \IBCS.
The complete list is:
$(SU(p,q), SU(p) \otimes SU(q))$, $(SO(p,2), SO(p))$,
$(Sp(2p,\IR), SU(p))$, $(SO^*(2p), SU(p))$, $(E_6, SO(10))$ and $(E_7, E_6)$,
where the first entry in the parentheses refers to the group $G$ and the
second one to the subgroup $H$. The lowest dimensional examples
have $D=6$, i.e. $(G, H)=(SO(2,2), SO(2))$ or $(SO^*(4), SU(2))$.

The actions \smod\ and \smods\ as well as the corresponding dilaton fields
are still invariant under the transformations \gtr. A convenient gauge choice
would be to set to zero all the parameters in $h_x$ except those
corresponding to the Cartan subalgebra of $H$ which cannot be gauged away due
to the existence of a non-trivial isotropy
(for discussions relevant to this point see also \nadual\grnonab).
That still leaves a number of parameters equal to
the rank of $H$ to be fixed among the remaining ones $\{a^{\a}\}$ and
$\{v^i\}$ (in the second set those with an $H$-Cartan subalgebra index are
inert under the remaining gauge transformations in \gtr).
As in the case of the WZW action \wzwc\ that follows directly from \unc\
through the limiting procedure described in the previous section, one can also
show that \gwzwc\ can be obtained from the usual gauged WZW action for
the coset $(G\otimes H)/H$ (the gauge group is the total subgroup generated
by $u^i+v^i$)
\eqn\gaa{S=k_G\ [\ I_0(h\inv t h_{x'} \bh) - I_0(h\inv \bh)\ ]
+k_H\ [\ I_0(h\inv h_x \bh)- I_0(h\inv \bh)\ ]\ ,}
via a similar procedure.
Besides the expansion formulae \te, \cartan\ and \car\
the following one
\eqn\expp{ \Tr( u_i  t_{\e} h_x u_j h_x\inv t_{\e}\inv ) = C_{ij}
+ \e\ (2 n + m m^t)_i{}^k C_{kj} + O(\e^2)\ , }
should also be used.

Let us also briefly discuss how possible $\a'\sim 1/\b$ corrections to our
$\s$-model \smods, \dilas\ can arise, by considering the effective action
for it. Again the connection with the original
$(G\otimes H)/H$ proves useful.
The effective action for the latter models is \TSEYT\BSeaction\
(we ignore possible field renormalizations since they give rise to non-local
terms in the final $\s$-model \TSEYTt)
\eqn\efact{\eqalign{S&=
[\ (k_G +g_G) I_0(h\inv t h_{x'} \bh) - (k_G +g_H) I_0(h\inv\bh)\ ]\cr
&+[\ (k_H + g_H) I_0(h\inv h_x \bh) - (k_H+ g_H) I_0(h\inv \bh)\  ]\ ,\cr} }
where all the definitions were given in \orismoi.
Then in the limit $\e\to 0$ this reduces to the effective action for our
$\ghc/H$ coset models
\eqn\efactc{\eqalign{\G= &(\b + g_G + g_H)\ I(h_x,A) +
{\a\ov 2\pi}\int_{\Sigma} d^2z\ [\
Da\cdot {\bar D}a + i (Da)^{\a} m_{\a}{}^i ({\bar D} h_x h_x\inv)_i \cr
&-2i \Tr( Dv {\bar D}h_x h_x\inv - v F_{z\z})\ ]
+(g_G-g_H) I_0(h\inv \bh)\ .\cr } }
The above action although local in the $h$-fields is not local
in the gauge fields $A$ and $\A$, due to the presence of the last term.
Of course in the limit of large $\b$ it reduces to the action given in \gwzwc.
Next, one is to solve for the gauge fields
via their equation of motion and identify the local part containing the
second derivative terms as
the final $\s$-model. We will not repeat this procedure here as it has been
discussed extensively in \BSeaction\TSEYTt\ and especially in \KST\ and
because the
final expressions for the $\s$-model are quite complicated.
We will only mention that for the metric and the dilaton there are $1/\b$
corrections in agreement with what is expected from the
different shiftings of $\b$ in the $F^2$ and the $T^2$ terms in the
expression of the stress tensor \cosetc.
For the antisymmetric tensor it was shown in \KST\ that there are
two natural prescriptions, consistent with gauge invariance, for extracting it.
The one which was called
`corrected' gives $1/\b$ corrections to the semiclassical result for the
antisymmetric tensor. However, the second prescription gives just the
semiclassical result. At this point we should mention that for the case of
the coset $E^c_2/U(1)$ the explicit
expressions containing all quantum $1/\b$ corrections were derived in \etc,
both in the axial and the vector gauging cases.
In the case of the axial gauging both of the prescriptions mentioned give the
semiclassical
result for the antisymmetric tensor. Being derived via a correlated limit from
the $3D$ charged black string \HOHO\SFET\ its conformal invariance has already
been checked up to two loops in conformal perturbation theory in \KST.

All of the above discussion is simplified in an unambiguous way if $H=G$.
In such cases, since $g_H=g_G$ the effective action \efactc\ is the same as
\gwzwc\ up to the shift $\b\to \b+g_G +g_H$ (see footnote 3), and therefore
there are no corrections at all to the $\s$-model action \smods\
and the dilaton \dilas, which further simplify since $m_{i\a}=0$.
In contrast to the case of a $G/G$ coset model where we can gauge out all
degrees of freedom, except those corresponding to the Cartan torus that
constitute a free field theory, in the case of a $G^c_h/G$ coset model
the resulting $\s$-model is a non-trivial one.
This fact receives more significance in view of the results that will follow
and specifically to the relation to non-abelian duality transformations
considered in the next section.

\newsec{ On non-abelian duality }

In this section we formulate the non-abelian duality transformations \nadual\
as a particular limiting case of a larger class of models where an anomalous
free symmetry of the target spacetime geometry is being gauged.
Moreover we show how the gauged WZW models corresponding to the cosets
$\ghc/H$, considered in the
previous section, can be thought of as non-abelian duality transformations on
the $H\otimes U(1)^{dim(G/H)}$ WZW model with respect to the vectorial action
of $H$.

\subsec { Non-abelian duality transformations and gauged WZW models }

In the usual Lagrangian formulation (for a Hamiltonian formulation see \GRV)
of non-abelian duality \nadual\grnonab\AABL\GRV\
one starts with an action $S(X)$ corresponding to some CFT,
denoted by $\cal A$, and then gauges a non-anomalous symmetry
corresponding to a group $H$ and adds a Lagrange multiplier term
(of course the same procedure works for any non-linear, not necessarily
conformal, $\s$-model with a global symmetry).
The total action reads
\eqn\sduul{ S(X,A,\A,v)=
S(X,A,\A) + {\a\ov \pi} \int_{\Sigma} d^2 z\ i\ \Tr(v F_{z\z} )\ .}
Each terms in the above action is invariant under gauge transformations
which for $A$, $\A$ and the $v^i$'s were given in \gtr.
The `matter' fields $X$'s transform in some representation of the gauge group
$H$ in such a way that the action $S(X,A,\A)$ is gauge invariant by itself.
In the path integral for $S(X,A,\A,v)$ integration over the Lagrange
multipliers $v^i$'s forces the
gauge fields to be pure gauges, i.e. $A=\del h h\inv$, $\A=\bd h h\inv$.
Choosing the gauge condition $h=I$ one recovers the original model with the
action $S(X)$. If instead
one integrates out the gauge fields one obtains the dual model.
In the case of non-abelian duality the original and the dual model do not
necessarily correspond to the same CFT \nadual\grnonab\AABL,
as in the case of the
abelian duality for compact groups \RV. Instead it has been conjectured
that non-abelian duality transformations interpolate between solutions of
different CFT's possibly related by an orbifold
construction \nadual\grnonab.

Next we will show how \sduul\ can be thought of as a limiting case of a
more general gauged theory.
Consider the direct product of the CFT's
$\cal A$ and the current algebra for $H$ at level $k_H$.
In order to gauge the $H$-symmetry the appropriate action is
\eqn\gds{{\cal S}(X,A,\A,h_x)= S(X,A,\A) + k_H\ I(h_x,A)\ ,}
where $I(h_x,A)$ is the gauged WZW action for $H/H$ as in \usual.
Let us consider the path integral for the above action in the limit of very
large $k_H$, or equivalently the limit $\e\to 0$ with $k_H={\a\ov 2\e}$
and $\a$ a finite constant.
In that limit the only contribution to the path integral from integrating
over all elements $h_x\in H$ comes from configurations close to the identity
element.\foot{We assumed that $h_x\in H$ belongs to an irreducible
representation. In that case all fixed points of the gauge
transformation (points $h_0\in H$ such that
$[h_0,u_i]=0,\ \forall\ u_i\in {\cal L}(H)$) are proportional to the identity
element according to Schur's first Lemma.}
For all other configurations the action in \gds\
becomes very large thus giving negligible contribution to the
path integral. If we parametrize
$h_x=I+ 2i\e\ v\cdot u$ then we can easily show that in the limit
$\e\to 0$ the action \gds\ becomes identical to the action \sduul\ and the
two theories that they describe become equivalent. Namely we have proved
\eqn\inee{
\int [dX][dh_x][dA][d\A] e^{-{\cal S}(X,A,\A,h_x) }\bigg|_{k_H\to \infty}
= \int [dX][dv][dA][d\A] e^{-S(X,A,\A,v)} \ .}
It is worth noticing that the Lagrange multiplier term in \sduul\
is nothing but the dimensionally reduced non-abelian Chern-Simons action
($v^iu_i$ is the third component of the $A$ field).\foot{Abelian versions
of that term were also derived by \WIT\ in showing that the gauged WZW action
for the coset model $SL(2,\IR)/\IR$ close to the $uv=1$ `singularity'
looks like a topological field theory and by \blau\ in the derivation,
from a Chern-Simons theory, of the Verlinde
formula that counts the number of conformal blocks of a rational CFT.}
It should be emphasized that the equivalence relation
\inee\ holds up to a total derivative which however is important when one
discusses global issues \RV\GR\grnonab\AABL.
In this paper we are only concerned with the
local (short distance) conformal properties of the models for which
the equivalence relation \inee\ (and also (4.4) below) is unambiguous.
Let us be more specific and consider for the CFT $\cal A$ the one corresponding
to the WZW for a group $G$ at level $k$. What we have proved is that
\eqn\ine{ {G_k \otimes H_{k_H} \ov H_{k + k_H}}\bigl|_{k_H\to \infty}
\qq \Longleftrightarrow  \qq
{\rm dual\ of}\ G\ {\rm with\ respect\ to}\ H\ ({\rm vector})\ .}
Our formulation of non-abelian duality as a limiting case of a larger class
of gauged models makes it possible to work out explicitly $\a'$-corrections
to the
semiclassical expressions. It is clear from the whole discussion in this
section that they can only arise, in an effective action approach, from
possible renormalizations in the action term $S(X,A,\A)$ in \sduul.
We can illustrate this briefly by considering the case of duality
transformations on a WZW model for a simple group $G$ with respect to its
subgroup $H$. In that case the effective action replacing \sduul\ is
\TSEYT\BSeaction\ (cf. \efactc)
\eqn\effd{\G= (k +g_G) I(g,A) + (g_G-g_H) I_0(h\inv\bh)
+{\a\ov \pi} \int_{\Sigma} d^2 z\ i\ \Tr(v F_{z\z} )\ ,}
where $g\in G$ is parametrized in terms of $x^{\m}$, $\m=1,2\dots, dim(G)$.
We will not present the derivation of the exact $\s$-model background fields
here. The method is identical to the one followed in \BSeaction\TSEYTt\KST\ for
the case of $G/H$ coset models with only some modifications. The result is
\eqn\sloc{S(x)=
 {1 \over \pi \a' } \int_{\Sigma} d^2 z \ ( G_{MN}+B_{MN})
\del x^M \bd x^N \ ,\qq x^M=\{ x^{\m}, v^i \}\ , \qq \a' = {2\ov k+g_G } \ ,}
where the metric is
\eqn\metri{ G_{MN}  = { G}_{MN }^{(s)}
-  2 b (  \ctV\inv \cM^t \cM\inv )_{ij} \cL^i_{(M} \cR^j_{N)}
- b (\ctV\inv )_{ij} \cL^i_M \cL^j_N -   b (\cV\inv )_{ij} \cR^i_M \cR^j_N \ ,}
with $b\equiv -{g_G - g_H \ov k + g_G}$, the antisymmetric tensor is
\eqn\ants{\eqalign{B_{MN}  =&{ B}_{MN }^{(s)} -  \
2b^2\bigl[ \ctV\inv \cM^t \cM\inv(\cM-\cM^t)\cV\inv \bigr]_{ij}
\cL^i_{[M} \cR^j_{N]} \cr
& - b^2 \bigl[ \ctV\inv(\cM-\cM^t)\ctV\inv \bigr]_{ij}  \cL^i_{[M} \cL^j_{N]}
- b^2 \bigl[ \cV\inv(\cM-\cM^t)\cV\inv \bigr]_{ij}  \cR^i_{[M} \cR^j_{N]}\ .} }
and the dilaton is
\eqn\dilaa{ \Phi = \ha  \ln \det \cV +\Phi_0  \ . }
The semiclassical expressions, in the limit $b\to 0$, are
\eqn\semicl{ {G}_{MN }^{(s)} ={ G}_{0MN }  -
 2 \cM\inv_{ij} \cL^i_{(M} \cR^j_{N)} \ , \quad  { B}_{MN }^{(s)}=B_{0MN} -
  2\cM\inv_{ij} \cL^i_{[M} \cR^j_{N]} \ , \quad \Phi^{(s)}= \ln \det \cM\ . }
${ G}_{0MN }$  and ${ B}_{0MN }$  are  the  original  WZW
(group space) couplings,
\eqn\orc{ G_{0MN} =\eta_{AB} L^A_M  L^B_N  \ , \qq
 3\del_{[K} B_{0MN]} = i L^A_KL^B_ML^C_N f_{ABC} \ ,}
and the various matrices are defined as
\eqn\oris{ \cM_{ij}=M_{ij} +  n_{ij}\ ,
\quad \cV=\cM\cM^t -b\ (\cM + \cM^t)\ ,
\quad \ctV=\cM^t \cM -b\ (\cM + \cM^t)\ ,}
where we have rescaled $v^i\to (k+g_G)/\a\ v^i$ and
\eqn\LR{ \cL^i_M \del x^M = L^i_{\m} \del x^{\m} + \del v^i\ ,
\qq \cR^i_M \bd x^M = R^i_{\m} \bd x^{\m} +  \bd v^i\ .}
Due to the gauge invariance we should fix $dim(H)$ of the parameters among
the $x^M$'s.
The remaining $dim(G)$ variables will be the string coordinates of the
dual space of $G$ under non-abelian duality with respect to a subgroup $H$.
Thinking of them as $H$-subgroup invariants one can determine their range of
values using group theoretic methods \BSglo.
If $H\neq G$ then there is generically no isometry and all of the parameters
to be gauge fixed can be chosen entirely within the group element $g\in G$.
Obviously if $H=G$ then since $b=0$ all the quantum corrections in \sloc\
drop out and we
deduce that the semiclassical results \semicl\ are also exact.
For such cases there is non-vanishing isometry and the
remarks of the paragraph just before \gaa, about a possibly convenient gauge
choice are equally applicable.
Let us note that the exact expression for the antisymmetric tensor \ants\
was obtained with the `corrected' prescription of \KST. As it was explained
also in section 3 there is a
second prescription that leads to the semiclassical result in \semicl\ for
the exact antisymmetric tensor.
Both prescriptions give the same metric and dilaton, are consistent
with the gauge invariance of \sloc\ (before gauge fixing) and
with results available from conformal perturbation theory \KST.
The $\s$-model corresponding to the case of $G=H=SU(2)$ was previously
considered in \grnonab.
What we have proved is that it is a limit of the $\s$-model for the coset
$SU(2)_{k_1} \otimes SU(2)_{k_2}/SU(2)_{k_1 +k_2}$ considered in \CRE\foot{
This coset was also considered in \BSthree\FRA\ for the case $k_1=k_2$
where the geometry of the resulting $\s$-model is easier to interpret.
Obviously, in this case correspondence with non-abelian duality cannot be
made.} and
that the explicit expressions found in \grnonab\ should be consistent with
conformal perturbation theory to all orders in $\a'$.

Since the equivalence relation \inee\ involves a {\it singular} limit the
final dual theory is not necessarily equivalent to the original corresponding
to the action $S(X)$ even though some quantities like the central
charge in the case of \ine\ are the same (equal to the central charge
of $G_k$).
In the case of duality transformations with respect to an abelian subgroup
we can avoid taken any singular limit.
This is the case because under the vector gauge
transformations \gtr\ any point in the group element $h_x=e^{i x\cdot u}$
is a fixed point, i.e. $\d x^i=0,\ \forall\ i$.
Then if in
$I(h_x,A)={1\ov \pi} \int d^2z\ (\ha \del x_i \bd x^i +i A_i \bd x^i -
i \A_i \del x^i)$ we shift $A_i\to A_i +i/4\ \del x_i$
and $\A_i\to \A_i -i/4\ \bd x_i$ (this is not a gauge transformation)
we can absorb the bilinear in the $x$'s term
and then \gds\ takes the form of \sduul\ with $\a= k_H$ and $v^i=x^i$.
The redefined $A$, $\A$ have the same transformation properties as the old
ones and they are the ones to be used in $S(X,A,\A)$.

\subsec{  Relation of $\ghc/H$ to non-abelian duality }

Let us uncover the relation of the gauged WZW model for the coset $\ghc/H$
that we considered in section 3 to the class of models one obtains
via non-abelian duality transformations
performed on the background corresponding to the WZW model for the direct
product group $H\otimes U(1)^l$. In particular we will consider non-abelian
duality transformations with respect to the group $H$ itself.
Since we do not
want to leave the coordinates corresponding to the factor $U(1)^l$ inert under
the duality transformation we embed the model
into a larger one for the group $G$ that contains $H$ as a subgroup.
In this way the gauge
transformations are as in \gtr, with the $a^{\a}$'s parametrizing
the original $U(1)$ factors whose number is obviously
restricted to be $l=dim(G/H)$.
The gauged invariant action we start with is (cf. \sduul)
\eqn\sdualg{S=\b\ I_0(h_x,A) + {\a \ov 2\pi}\int_{\Sigma} d^2 z\
[\ Da\cdot {\bar D}a + 2 i\ \Tr(v F_{z\z})\ ]\ ,}
where all the necessary definitions are given by \orismoi.
Integrating over the gauge fields one obtains the dual action
\eqn\dual{\eqalign{&\tilde{S}_{\rm dual}= \b\ \bigl \{\ I_0(h_x)
+  {1\ov 2\pi} \int_{\Sigma} d^2 z\ [\ \del \a \cdot \bd a \cr
&-2 (L^i_{\m} \del x^{\m} + \del v^i + \ha m^i{}_{\a} \del a^{\a})
(M + n + \ha mm^t)\inv_{ij} (R^j_{\n} \bd x^{\n}
+ \bd v^j - \ha m^j{}_{\b} \bd a^{\b}) ] \bigr \}\ . \cr }}
The induced dilaton is
\eqn\dilad{\tilde{\Phi}= \ln \det \bl (M + n +\ha m m^t \br ) + \Phi_0\ .}
A quick inspection (and taking into account the rescaling as described in
footnote 5) shows that if in \dual\ and \dilad\ we send
$\b\to -\b$ and $h_x\to h_x\inv$ we obtain precisely \smods\ and \dilas.
In fact correspondence can also be made for the
`original' models one obtains by integrating over the Lagrange multipliers
$v^i$'s in \gwzwc\ and \sdualg. For the latter case the result is, as we have
already mentioned, the WZW model for $H\otimes U(1)^{dim(G/H)}$
\eqn\wzwd{S=\b\ I_0(h_x) + {\a\ov 2\pi}
\int_{\Sigma} d^2 z\ \del a\cdot \bd a\ .}
Varying \gwzwc\ with respect to the $v^i$'s and using the fact that
\eqn\facto{ F_{z\z}= D(\A-\tA)\ ,\qq A=\del h h\inv\ ,\qq \tA=\bd h h\inv}
we obtain the equation
\eqn\factt{{\bar D} h_x h_x\inv + \A - \tA=0\ .}
Choosing the gauge fixing condition $h=I$ one obtains $A=\tA=0$ and
$\A=-h_x\inv \bd h_x$. After a little algebra \gwzwc\ becomes
\eqn\wzwa{S=-\b\ I_0(h_x\inv) + {\a\ov 2\pi}
\int_{\Sigma} d^2 z\ \del a\cdot \bd a\ ,}
thus revealing the same relationship between the two models we have already
uncovered by comparing \smods\ and \dilas\ to \dual\ and \dilad.  It appears
as if the central charge corresponding to the model \wzwa\
is not the same as in
\cosetc, i.e. `naively' it is $c=dim(G) + g_H dim(H)/(\b-g_H)$.
However, in order to correctly compute it, one has to take into account the
non-trivial
Jacobian arising from changing variables from $(A,\A)\to (h,\bh)$ in the path
integral
functional before the gauge fixing condition, $h=I$, is imposed.
The Jacobian regularized in a gauged invariant way \SCH\ gives a factor
\eqn\ano{{\cal J}=e^{2g_H I_0(h_x\inv)}\ \bl(\det\ \del\bd\br)^{dim(H)} }
which shifts the value of $\b\to \b+ 2g_H$ in \wzwa.
Thus the central charge is given by
$$c= {-(\b +2 g_H)dim(H)\ov -(\b +2 g_H) + g_H} + dim(G/H) + 2 dim(H)
-2 dim(H) \ ,$$
which produces the correct central charge as given by \cosetc.
In the previous expression the second term is due to the $a^{\a}$'s and the
last two terms due to the contributions of
the $v^i$'s and the factor $\det (\del \bd)^{dim(H)}$ in \ano\
(they cancel because each factor corresponds to $dim(H)$
independent $(b,c)$ systems of conformal weight $(0,1)$ but of opposite
statistics). Therefore we have proved the relation
\eqn\rela{ \ghc/H\qq  \Longleftrightarrow \qq
{\rm dual\ of}\ H\otimes U(1)^{dim(G/H)}\ {\rm with\ respect\ to}\ H\
({\rm vector})\ ,}
with the specific relation between the central extension parameters
that we have already mentioned. The equivalence relation \rela\ is very
similar to \ine. In fact it seems that it can be directly deduced from it,
without having to go through the explicit computations of this subsection.
However, this is not true because there is a double correlated limit to be
taken in \ine\ instead of the simple one $k_H\to \infty$.
Let us also stress that \rela\ is non-trivial in the sense that the
actions describing the two sides of it, namely \gwzwc\ and \sdualg\ appear
quite different from one another.

\newsec{ Concluding remarks and discussion }

In this paper we have been studying limiting cases of WZW and gauged WZW models
based on simple groups both at the algebraic and the $\s$-model action level.
The resulting models after the limiting procedure is performed are WZW and
gauged WZW models based on a certain class of non-semi-simple groups.
We have seen that there is an intimate connection between these models
as well as non-abelian quotient models, and the models one obtains via
non-abelian duality transformations. This correspondence facilitates a lot
the systematic computation of the $\a'$-corrections to the
semiclassical results of the non-abelian duality transformations,
as we have already seen. A more difficult question to be answered
is how the spectrum of the dual theory is related to that of the original one.
If the original theory is the WZW model for $G_k$ then the equivalence
relation \ine\ is potentially useful.

A motivation for studying gauged WZW models based on non-semi-simple
groups was that according to the results of \etc\ for the coset $E^c_2/U(1)$,
the $\s$-model background fields corresponding to the $\ghc/H$ cosets might be
possible to be mapped to non-singular ones, even though themselves may
have curvature singularities.
In other words, even though the original background for the coset
$(G\otimes H)/H$ cannot be dual to a non-singular one (for instance,
the charged $3D$ black string is dual to the neutral one and vice versa \SANT,
but both backgrounds have curvature singularities), the one corresponding
to the `contracted' coset $\ghc/H$ can.
This would have been a possible and quite general way String
theory deals with gravitational singularities
(although the latter are not peculiar from a CFT point of view \WIT\GIN).
We partially succeeded towards this goal.
In the case of an abelian subgroup $H$
of a general group $G$ it is shown in Appendix A that it
is possible to map the curved singular backgrounds to flat spacetimes with
constant antisymmetric tensor and dilaton fields
under an abelian duality transformation. The reason that for the non-abelian
case we were not able to make a similar statement is that, in view of the
connection to non-abelian duality transformations, such an `inverse'
transformation is not known how to be performed and as we have already
mentioned the symmetry of the dual background is much less than that of
the original one. Thus finding a way to define the `inverse' non-abelian
duality transformation will give a definite answer to whether
or not such a desired mapping is possible.

A natural question is whether or not the construction of \ORS\ and the similar
one in Appendix B exhaust all possible non-semi-simple algebras with a
sensible action description.
It will be desirable for instance to find the action
corresponding to the CFT whose Virasoro construction is given by \virstr,
using possibly the actions of \TSEma\BCHma. We should also mention that in
\noumo\ a formalism for constructing WZW models based on non-semi-simple
groups that generically give rise to non-integer values for the central charge
was proposed. For the models of \ORS\ this formalism corresponds to a shift
of the constant $\b$ defined in \conta\ and it gives no new results
(the essential reason for that is the fact that $\Om^{T_iT_j}=0$).
It will be of interest to search for models where the results of \noumo\
are applicable in a non-trivial way.

\bs\bs

\centerline { \bf Acknowledgments }

I would like to thank E. Kiritsis and A.A. Tseytlin for useful discussions
and comments.

\bs\bs

\centerline {\bf Note added }

After the completion of this work we received ref. \figsonia\ which proves that
there are no Sugawara constructions (with respect to which all currents are
primary fields with conformal weight one)
based on non-semi-simple algebras that give
rise to non-integer values for the central charge along the lines of \noumo.

We would like also to suggest a possible explanation for the origin of a
problem with non-abelian duality that was noted in \GRV.
In that paper the
three dimensional symmetry algebra of the spatial part of the metric in which
the non-abelian duality transformation is performed admits no Sugawara
construction. One can easily prove that by a direct computation or by using
the result of \figsonia\ according to which the only such three dimensional
argebras are $su(2)$, $sl(2)$ and $u(1)^3$.
That will give rise to a conformal anomaly \GRV\
when we perform the non-abelian duality transformation since the
assumption that the currents coupled to the gauged fields in the action
scale with dimension one is not satisfied at the quantum level.
A better and detailed understanding of it is important but beyond the scope
of this note.
I would like to thank R. Ricci for motivating me to think about this problem.

\vfill\eject

\appendix A {Axial gauging }

In this Appendix we consider the case of the axial gauging. This is anomaly
free when the gauge group is abelian, i.e. isomorphic to $U(1)^d$. We will
show how the action for the gauged WZW models $\ghc/H$ in the axial gauging
can be obtained directly from the action of the gauged WZW models (in the axial
gauging as well) for the $G\otimes U(1)^d/U(1)^d$ through a limiting procedure
and also how duality transformations can be used to map the final $\s$-models
(with curvature singularities in general) to {\it flat spacetimes}
with constant antisymmetric tensor and dilaton fields.
For the former models the group element is parametrized as in
\param\ with $h_x=e^{i T\cdot x}$. In the axial gauging the corresponding
action is given by (see for instance \WIT\BSthree\GIN)
\eqn\axial{S=  I_0(h\inv g \bh) - I_0(h\bh)\ ,}
where $h$, $\bh$ are groups elements in $H\simeq U(1)^d$. Defining
$A=\del h h\inv$ and $\A=\bd \bh \bh\inv$ and shifting $\A_i\to \A_i -i\bd x_i$
(this will, among other things, effectively change the coefficients
$\b_i \to - \b_i$ below)
we obtain with the help of \lj\ and \dggi\ the following action
\eqn\axia{\eqalign{S_{\rm axial}= &{1\ov 2\pi} \int_{\Sigma} d^2z\ \bigl \{\
\a \del a\cdot \bd a - \b_i \del x_i \bd x_i
+ i A_i\ [\ \a(2 \bd v_i -m_{i\a} \bd a^{\a}) -2 \b_i \bd x_i\ ] \cr
&- i \A_i\ [\ \a (2 \del v_i + m_{i\a} \del a^{\a}) + 2 \b_i \del x_i\ ]
+ A_i\ [\ \a (m m^t)_{ij} + 4 \b_i \d_{ij}\ ]\ \A_j\ \bigr \}\ ,\cr } }
where summation oven repeated indices is implied. This action is invariant
under the infinitesimal gauge transformations
(in fact \axia\ can be cast in a similar
to \gwzwc\ form, i.e. as the first two terms with covariant derivatives
replacing the ordinary ones plus the Lagrance multiplier term)
\eqn\gtra{\d x_i= 2 \e_i\ , \quad \d a^{\a} = m^{\a}{}_j \e^j\ ,
\quad \d v_i =0\ ,\quad \d A_i = i\del \e_i\ ,\quad  \d\A_i = i \bd \e_i\ .}

Let us next consider the action for the axially gauged WZW models
$(G \otimes U(1)^d)/U(1)^d$
\eqn\aca{\eqalign{S=& k\ I_0(g)
+{k\ov \pi} \int_{\Sigma} d^2 z\ [\ A\bd g g\inv
- \A g\inv \del g +A g \A g\inv + A\A \ ] \cr
& + {1\ov \pi} \int_{\Sigma} d^2 z\ k_i\ [\ \ha \del \g_i \bd \g_i
+ i A_i \bd \g_i - i \A \del \g_i +2 A_i \A_i \ ]\ ,\cr } }
where the $\g_i$'s parametrize the $U(1)^d$ factor in $G \otimes U(1)^d$.
The above action is invariant under the infinitesimal
axial gauge transformations
\eqn\gttr{ \d g=\{g,i\e\}\ ,\quad \d \g_i= 2 \e_i\ ,\quad \d A_i=i \del \e_i\ ,
\quad \d \A_i = -i \bd \e_i \ .}
The group element $g\in G$ is parametrized as
\eqn\parr{g= e^{i\sqrt{2\e}\ a\cdot R}\ e^{2 \e i v \cdot u}\ .}
Using the expansions in powers of $\e$ (very similar to the ones in \cartan)
\eqn\expa{\eqalign{ & g\inv \del g= i\sqrt{2\e}\ \del a\cdot R +
i \e\ [\ 2 \del v\cdot u  - \del a^{\a} m_{\a}{}^i u_i
+ \a^{\a} \del a^{\b} S_{\a\b}{}^{\g} R_{\g}\ ] + O(\e^2) \cr
& \bd g g\inv = i \sqrt{2\e}\ \bd a\cdot R + i \e\ [\ 2 \bd v\cdot u
+ \bd a^{\a} m_{\a}{}^i u_i
- \a^{\a} \bd a^{\b} S_{\a\b}{}^{\g} R_{\g}\ ] + O(\e^2) \cr
& \Tr(A g \A g\inv) = A_i\A_i + \e\ A_i(m m^t)_{ij} \A_j + O(\e^2)\cr
& I_0(g) = {\e\ov \pi} \int_{\Sigma} d^2 z\ \del a\cdot \bd a + O(\e^{3/2})
\ ,\cr } }
choosing the gauge fixing condition $\g_i=0\ ,i=1,2,\dots, d$ and letting
\eqn\letting{ k={\a\ov 2\e}\ ,\qq k_i=\b_i -{\a\ov 2\e} }
we find that the action \aca\ in the limit $\e\to 0$ becomes the action \axia\
in the gauge $x_i=0\ ,i=1,2,\dots, d$.
Notice that in \axia\ the $v^i$'s enter as Lagrange multipliers
(up to an important total derivative \RV\GR) for
abelian duality transformations in agreement with the general discussion in
section 4. Moreover, since the duality transformations are abelian it is
apparent that the final $\s$-model one obtains by integrating out the
gauge fields will have $dim(H)$ Killing vectors along the $v^i$-directions,
i.e. it will be
invariant under the constant shifts $v^i\to v^i + \e^i$ (in the non-abelian
case this symmetry gets replaced by a non-local one \grnonab).
By gauging this symmetry or in other words by performing the inverse duality
transformation one obtains the background, in {\it flat spacetime},
described by \axia\ after setting the gauge fields to zero
(for a nice proof that two successive duality transformations corresponding
to the same {\it abelian} isometries lead to the original model see \grnonab).
This is generalization of a similar statement made in \etc\
at the $\s$-model level (after integrating over the gauge fields)
for the case of $ E^c_2/U(1)$ which was shown to be related to the
$D=3$ black string $SL(2,\IR) \otimes \IR/\IR$.

\appendix B {Contraction of $G \otimes H \otimes H'$ }

In this Appendix we construct new WZW models based on non-semi-simple groups
via a similar to the section 2 limiting procedure.
Let us consider the WZW model for $G \otimes H \otimes H'$ where $G$, $H$ and
$H'$ ($H$, $H'$ are assumed isomorphic) are groups and as before $G$ should
contain a subgroup isomorphic to $H$ and $H'$. We choose a basis for the
currents:
$\hat g=\{u_i,R_{\a}\}$, $\hat h=\{v^+_i, v^-_i\}$,
where $i=1,2,\dots, dim(H)$ and
$\a=1,2,\dots, dim(G/H)$. Above $\hat g$ and $\hat h$
belong to the current algebras associated with the WZW models for the groups
$G$ and $H\otimes H'$ respectively. The OPE's for $\hat g$ are given by the
corresponding ones in \opea, whereas those for $\hat h$ are
\eqn\opeb{v^{\pm}_i v^{\pm}_j \sim {i f_{ij}{}^k v^+_k \ov z-w}
+ {k^+\ \eta_{ij} \ov (z-w)^2}\ ,\qq
v^+_i v^-_j \sim {i f_{ij}{}^k v^-_k \ov z-w} + {k^-\ \eta_{ij}\ov (z-w)^2}\ .}
The corresponding energy momentum tensor and the associated central charge are
\eqn\virb{\eqalign{&T={:u^2 + R^2:\ov 2(k_G+2 g_G)} + {:v_1^2:\ov 2(k_1+g_H)} +
{:v_2^2:\ov 2(k_2 +g_H)}  \cr
&c={k_G\ dim(G)\ov k_G + g_G } + {k_1\ dim(H)\ov k_1 + g_H}
+ {k_2\ dim(H)\ov k_2 + g_H}\ ,\cr} }
where the currents $(v_1)_i=(v^+_i + v^-_i)/2$ and
$(v_2)_i=(v^+_i - v^-_i)/2$ generate two commuting copies
of the Kac-Moody algebra with levels $k_1=(k^+ + k^- )/2$ and
$k_2=(k^+ - k^-)/2$ respectively. Next we define
\eqn\contb{\eqalign{&T_i=u_i + v^+_i\ ,\qq F_i=\e(u_i-v^+_i)\ ,
\qq P_{\a}= \sqrt{2\e}\ R_{\a}\ ,
\qq S_i= \sqrt{2\e}\ v^-_i - {\g\ov \a}\ F_i \cr
&k_G=\ha (\b +\a/\e)\ ,\qq k^+ = \ha (\b - \a/\e)\ ,
\qq k^- = {1\ov \sqrt{2\e}} \g\ ,\cr} }
and take the singular limit $\e\to 0$. In this limit we discover a new current
algebra {\it not} equivalent to the original one because the transformation
\contb\ is not invertible in that limit. The OPE's of the new current algebra
one obtains this way are given by \opeac\ and the additional ones
\eqn\opebc{
S_i S_j \sim {-i f_{ij}{}^k F_k \ov z-w} + {-a\ \eta_{ij}\ov (z-w)^2}
\ ,\qq T_i S_j \sim {i f_{ij}{}^k S_k \ov z-w} \ .}
We see that the current generators $S_i$ although they have a different index
structure are very similar to the $P_{\a}$'s and that the constant $\g$ drops
out completely.
The corresponding energy momentum tensor and the associated central charge are
\eqn\virbc{\eqalign{&T={:P^2 - S^2 +2 FT :\ov 2\a}
- {\b + g_G +2 g_H \ov 2 \a^2}\ :F^2: \cr
&c= dim(G) + 2 dim(H)\ .\cr} }
The OPE's in \opebc\ define the quadratic form
\eqn\quadb{\Om_{AB}=\bordermatrix{ & P_{\b} & T_j & S_j & F_j\cr
P_{\a} & \a/\b\ \eta_{\a\b} & 0 & 0 & 0\cr
T_i & 0 & \eta_{ij} & 0 & \a/\b\ \eta_{ij} \cr
S_i & 0 & 0 & -\a/\b\ \eta_{ij} & 0 \cr
F_i & 0 & \a/\b\ \eta_{ij} & 0 & 0  \cr }\ ,}
which by construction shares all three properties that are necessary if it is
to be used to write down a WZW action (it is symmetric, a group
invariant and invertible). Sparing the details we will give the final
expression for such an action. In the parametrization where
\eqn\pamamm{g= e^{ib\cdot S} e^{i a\cdot P} e^{i v\cdot F} h_x }
the WZW action, whose symmetry current algebra is given in \opeac, \opebc,
reads
\eqn\wzwcc{S(g)= \b\ I_0(h_x) + {\a\ov 2\pi} \int_{\Sigma} d^2z\
[\ \del a\cdot \bd a  - \del b \cdot \bd b + (2 \del v_i
+ m_{i \a} \del a^{\a} - b_{ij} \del b^j) R^i_{\m} \bd x^{\m}\ ]\ ,}
where $b_{ij}=f_{ikj} b^k$. Similarly to \wzwc\ the action \wzwcc\
describes string backgrounds in $dim(G) + 2 dim(H) $ spacetime dimensions
with $dim(H)$ null Killing vectors associated with the coordinates
$v^i$. The coordinates $b^i$ and $a^{\a}$ enter the action in a similar way
although they have a different index structure. If $H$ is abelian the theory
becomes equivalent to $\ghc \otimes \IR^{dim(H)}$.
It is better to analytically continue $S_i\to i S_i$. In that case and for
$G$, $H$ compact groups ($\a, \b>0$)
the signature of the spacetime of \wzwcc\ has
$dim(G) + dim(H)$ positive and $dim(H)$ negative entries.

\appendix C { Plane wave solutions }

Let us, in \wzwc, consider the case of $G=SO(d+2)$ and $H=SO(2)$. In this case
there is only one timelike coordinate. Also to agree with widely accepted
conventions in the literature we use the symbol $u$ for the single parameter
in $h_x\in SO(2)$. The
invariant subgroup $SO(d)$ after the contraction gives rise to $d(d-1)/2$
decoupled free fields in the action \wzwc\ which we shall ignore.
After a shifting in $v$, to absorb the $\del u\bd u$ term, and a rescaling
we obtain the action
\eqn\acb{S={\b\ov 2\pi} \int_{\Sigma} d^2 z\
[\ 2\del v \bd u + \del a^i_{\a} \bd a^i_{\a}
+ \e_{\a\b} \del a^i_{\a} a^i_{\b} \bd u\ ]\ ,}
where $\a,\b=1,2$ and $i=1,2, \dots , d$ and as usual summation over repeated
indices  is implied. Of course for $d=1$ this is the
result of \NW\ and the corresponding CFT is the current algebra for $E^c_2$.
Also notice that \acb\ is {\it not} the action for
just the direct product of $d$ $E^c_2$ models.
A straightforward extension of the change of variables used in \NW\
\eqn\acc{a^i_1= x^i_1 + x^i_2 \cos u\ ,\quad a^i_2 = x^i_2 \sin u\ ,\quad
v\to v+ \ha x^i_1 x^i_2 \sin u\ ,}
gives the final form of the action
\eqn\acd{S={\b\ov 2\pi} \int_{\Sigma} d^2 z\
[\  2 \del v\bd u + \del x^i_1 \bd x^i_1
+\del x^i_2 \bd x^i_2 + 2 \cos u\ \del x^i_1 \bd x^i_2\ ]\ .}
This belongs to a class of plane wave-type exact string solutions with a
covariantly constant null Killing vector that have been discussed extensively
in the literature (see for instance \Brinkman\SANTA).
However, what is important
here is that there is also a CFT description for \acd\ given by the
current algebra for $SO^*(d+2)_{so(2)}^c$, where the star implies that we
neglect the $d(d-1)/2$ free decoupled fields, as we have already stated.
For completeness we give
the corresponding OPE's (cf. \opeac)
\eqn\ace{\eqalign{& J P^i_{\a} \sim {i\e_{\a\b} P^i_{\b} \ov z-w} \ ,\qq
J F \sim {1\ov (z-w)^2}  \cr
& P^i_{\a} P^j_{\b} \sim {\d^{ij} \e_{\a\b} F\ov z-w}
+ {\d^{ij} \d_{\a\b} \ov (z-w)^2}\ ,\cr } }
and the Virasoro algebra stress tensor and central charge
\eqn\acf{T=\ha :\bl( P^i_{\a} P^i_{\a} +2 JF +F^2\br):\ ,\qq c=2(d+1)\ .}

\listrefs
\end